\documentclass[12pt,preprint]{aastex}

\slugcomment{Accepted for Publication in {\it The Astrophysical Journal}}

\shorttitle{Non-Thermal X-ray Emission from the Supernova Remnant G266.2$-$1.2}
\shortauthors{Pannuti et al.}

\begin{document}


\title{Non-Thermal X-ray Emission from the Northwestern Rim of the Galactic 
Supernova Remnant G266.2$-$1.2 (RX J0852.0$-$4622)}


\author{Thomas G. Pannuti\altaffilmark{1}, Glenn E.  Allen\altaffilmark{2}, Miroslav D. 
Filipovi\'{c}\altaffilmark{3}, Ain De Horta\altaffilmark{3}, Milorad Stupar\altaffilmark{4,5}
and Rashika Agrawal\altaffilmark{1,6}}


\altaffiltext{1}{Space Science Center, Department of Earth and Space Sciences, 
235 Martindale Drive, Morehead State University, Morehead, KY 40351; 
t.pannuti@moreheadstate.edu}
\altaffiltext{2}{Massachusetts Institute of Technology, Kavli Institute for Astrophysics
and Space Research, 77 Massachusetts Avenue, NE80-6025,
Cambridge, MA 02139; gea@space.mit.edu}
\altaffiltext{3}{University of Western Sydney, Locked Bag 1797, Penrith South DC,
NSW 1797, Australia; m.filipovic@uws.edu.au, a.dehorta@uws.edu.au}
\altaffiltext{4}{Department of Physics, Macquarie University, Sydney, NSW 2109, Australia;
mstupar@ics.mq.edu.au}
\altaffiltext{5}{Anglo-Australian Observatory, PO Box 296, Epping, NSW 1710, Australia}
\altaffiltext{6}{Department of Computer \& Electrical Engineering Technology \&
Information Systems and Technology, Indiana University - Purdue University Fort Wayne, 
2101 East Coliseum Blvd., Ft. Wayne, Indiana, 46805-1499; agar01@ipfw.edu}


\begin{abstract}
We present a detailed spatially-resolved spectroscopic analysis of two
X-ray observations (with a total integration time of 73280 seconds) made of
the luminous northwestern rim complex of the Galactic supernova remnant (SNR)
G266.2$-$1.2 (RX~J0852.0$-$4622) with the {\it Chandra} X-ray
Observatory. G266.2$-$1.2 is a member of a class of Galactic SNRs
which feature X-ray spectra dominated by non-thermal emission: in the cases of
these SNRs, the emission is believed to have a synchrotron origin and studies of 
the X-ray spectra of these SNRs can lend insights into how SNRs accelerate
cosmic-ray particles. The {\it Chandra} observations have clearly revealed fine 
structure in this rim complex (including a remarkably well-defined leading 
shock) and the spectra of these features are dominated by non-thermal emission.
We have measured the length scales of the upstream structures 
at eight positions along the rim and derive lengths of 0.02-0.08 pc (assuming a 
distance of 750 pc to G266.2$-$1.2). We have also extracted spectra from seven 
regions in the rim complex (as sampled by the ACIS-S2, -S3 and -S4 chips) 
and fit these spectra with such models as a simple power law as well as the 
synchrotron models SRCUT and SRESC. We have constrained our fits to the 
latter two models using estimates for the flux densities of these filaments 
at 1 GHz as determined from radio observations of this rim complex made with 
the Australia Telescope Compact Array (ATCA). Statistically-acceptable fits 
to all seven regions are derived using each model: differences in 
the fit parameters (such as photon index and cutoff frequency) are seen in the 
different regions, which may indicate variations in shock conditions and the maximum 
energies of the cosmic-ray electrons accelerated at each region. 
Finally, we estimate the maximum energy of 
cosmic-ray electrons accelerated along this rim complex to be approximately 
40 TeV (corresponding to one of the regions of the leading shock structure
assuming a magnetic field strength of 10 $\mu$G). We include a summary of 
estimated maximum energies for both Galactic SNRs as well as SNRs located in the
Large Magellanic Cloud. Like these other SNRs, it does not appear that 
G266.2$-$1.2 is currently accelerating electrons to the knee energy
($\sim$3000 TeV) of the cosmic-ray spectrum. This result is not surprising, 
as there is some evidence that loss mechanisms which are not important for 
the accelerated cosmic-ray nucleons at energies just below the knee might 
cut off electron acceleration. 
\end{abstract}

\keywords{acceleration of particles -- cosmic rays -- supernova remnants
-- X-rays: individual (G266.2-1.2)}

\section{Introduction\label{IntroductionSection}}

Non-thermal X-ray emission has been detected from a steadily growing number of
supernova remnants (SNRs) located in the Galaxy. Thanks to significant advances in 
the angular resolution capabilities of modern X-ray observatories, important progress has been
made in localizing the origin of this particular type of emission. Such localization has revealed 
that this emission is not associated with any neutron stars observed in projection toward these
SNRs (which may or may not be physically associated with the SNRs themselves): also, this
localization is essential for identifying the radiation mechanism responsible for this emission. In
the case of six SNRs -- namely SN 1006 \citep{Koyama95, Allen01, Dyer01, Dyer04, Vinketal03,
Rothenflug04,Bamba08}, G1.9$+$0.3 \citep{Reynolds08,Ksenofontov10},  G28.6$-$0.1
\citep{Bamba01,Koyama01,Ueno03}, G266.2$-$1.2 \citep{Tsunemi00, Slane01a, Slane01b,
Bamba05a, Iyudin05}, G330.2$+$1.0 \citep{Torii06, Park09} and G347.3$-$0.5 
\citep{Koyama97, Slane99, Pannuti03,CassamChenai04, Lazendic04, Hiraga05, Tanaka08} 
-- the detected X-ray emission is predominately non-thermal. For a subset of these SNRs 
(SN 1006, G1.9$+$0.3, G266.2$-$1.2, G330.2$+$1.0 and G347.3$-$0.5) the 
non-thermal X-ray emission is localized to narrow filaments of emission located on the leading
edge of the SNR. Other Galactic SNRs -- such as Cas A, Kepler, Tycho and RCW 86 --
feature significant amounts of non-thermal X-ray emission in addition to the typical 
thermal X-ray emission detected from SNRs \citep{Holt94, Allen97, Hwang00, Borkowski01,
Vink03, Bamba05b}.
In the cases of these particular SNRs, non-thermal
X-ray emitting filaments are observed surrounding regions of thermal-dominated X-ray emission.
In fact,  in the case of Cas A (and likely also to be the case for other SNRs as well), filaments of
non-thermal emission are also detected toward the central regions of the SNR: such filaments 
may be physically located on the edge of the SNR but seen toward the center due to projection
effects or they may actually be located
in the interior of the SNR. We note that Cas A also exhibits evidence for an unresolved component
to its non-thermal X-ray emission: evidence for the existence of this component includes 
measurements of the total flux from the sum of these filaments (including both filaments located
along the edge of Cas A as well as filaments seen toward the interior) can only account for 
one-third of the non-thermal X-ray emission detected from the SNR by other X-ray observatories.
Complementary {\it XMM-Newton} images of Cas A at higher X-ray energies (corresponding to
approximately 10 keV)  suggest that this unresolved component may be associated with the
reverse shock or contact discontinuity (in other words, a region located inside the SNR). In fact,
in an analysis of {\it Suzaku} observations of Cas A, \citet{Maeda09} found that the peak of
hard X-ray emission (for energies between 11 and 14 keV) possibly coincides with the location 
of an observed peak of TeV emission from the SNR: those authors suggest that this finding
indicates that both high-energy hadrons as well as leptons can be accelerated in the reverse
shock in an SNR, since the TeV peak should have a hadronic origin (see, for example,
\citet{Vink03}). 
\par
In the cases of those
SNRs which feature X-ray spectra dominated by non-thermal X-ray emission, a diffuse 
component of X-ray emission of this type is usually not detected. Based on 
data from pointed observations made 
with the Advanced Satellite for Cosmology and Astrophysics ($\it{ASCA}$) as well as 
data from the $\it{ASCA}$ Galactic Plane Survey \citep{Sugizaki99}, 
\citet{Bamba03} identified three new Galactic candidate X-ray SNRs 
(G11.0$+$0.0, G25.5$+$0.0 and G26.6$-$0.1) which may also belong to this 
class and \citet{Yamaguchi04} identified yet another possible member, G32.4$+$0.1. 
\citet{Bamba03} also estimated that approximately 20 Galactic SNRs 
with X-ray spectra dominated by non-thermal emission may be luminous enough to 
have been detected by the $\it{ASCA}$ Galactic Plane survey. Therefore, it
is reasonable to expect that even more SNRs of this type will be detected and 
identified as the sensitivities and imaging capabilities of X-ray 
observatories continue to improve and more X-ray observations of the
Galactic Plane are conducted.
\par 
Synchrotron radiation is a commonly accepted mechanism for the origin of the observed 
non-thermal X-ray emission detected from SNRs. In the case of a synchrotron origin, the
observed radiation is being produced by highly-relativisitic 
cosmic-ray electrons accelerated along the expanding shock fronts of these 
SNRs \citep{Reynolds96, Reynolds98, Keohane98, Reynolds99} and localized to the
observed filaments. Other processes besides synchrotron radiation -- namely inverse Compton
scattering of cosmic microwave background photons and non-thermal
bremsstrahlung -- have also been proposed to account for the observed
non-thermal X-ray emission from SNRs, but these two other proposed processes
both have difficulties in adequately modeling this emission \citep{Reynolds96, Rho02}.
A non-thermal bremsstrahlung origin for the emission has been ruled out 
because the same electron population responsible for such emission should
also excite emission lines in the X-ray spectra which are not observed. In the
case of a possible inverse Compton scattering origin, the predicted slope of the X-ray 
spectrum produced by this process is much flatter than the slopes of the detected
X-ray spectra. Additional evidence for a synchrotron origin of this observed X-ray
emission stems from qualitatively consistent simple modeling of synchrotron emission
from SNRs over broad ranges of photon energies, namely the radio through the X-ray
\citep{Reynolds99,Hendrick01}. Debate still persists in regards to the diffuse non-thermal 
X-ray emission detected from Cas A: some authors have argued that this emission has a
synchrotron
origin \citep{Helder08,Vink08} while others suggest a non-thermal bremsstrahlung origin
\citep{Allen08a}. We note that morphological comparisons between radio and
X-ray images of SNRs have also been made as part of studies of X-ray synchrotron
emission from these sources. These comparisons assume that the radio emission from
an SNR has a synchrotron origin and have concentrated on higher X-ray energies to
avoid confusion with thermal emission from the SNR at lower X-ray energies. The
results of these morphological comparisons have been mixed. A robust correspondence
between X-ray and radio emission along the northeastern rim of SN 1006 has been
noted \citep{Winkler97,Long03} as well as a broader overlap between X-ray and 
radio emission from the northwestern rim of G266.2$-$1.2 \citep{Stupar05}: these results
support a synchrotron origin for the non-thermal X-ray emission detected from these SNRs.
However, in the case of G1.9$+$0.3 an anti-correlation between X-ray and radio emission 
is observed: the northern rim is  a strong radio source but weak in X-ray while the eastern
and western rims are strong in X-ray but weak in radio \citep{Reynolds08}. Finally,
X-ray and radio emission from the SNRs G28.6$-$0.1. G330.2$+$1.0 and G347.3$-$0.5
do not appear to be strongly correlated \citep{Ueno03, Lazendic04, Park09}, but such 
comparisons are complicated by the fact that each of these SNRs are weak radio emitters. 
In Table \ref{SummaryObsSNRsTable}, we present 
a summary of observations that have been made with the {\it Chandra} X-ray Observatory 
\citep{Weisskopf02} of the six Galactic SNRs mentioned previously with X-ray spectra 
dominated by non-thermal emission. These studies have exploited
the high angular resolution ($\sim$ 1$\arcsec$) capabilities of {\it Chandra} 
to image the leading filaments located in the X-ray luminous rims of these SNRs. 
\par
A synchrotron origin for the observed non-thermal X-ray emission from the 
rims of these SNRs is of considerable interest to the cosmic-ray astrophysics 
community because such emission is consistent with models of synchrotron 
radiation from cosmic-ray electrons accelerated along the forward shock of SNRs 
\citep{Ellison01,vanderswaluw04,Huang07,Moraitis07,Ogasawara07, Katz08, Ellison08,
Berezhko08, Telezhinsky09, Morlino09}. By thoroughly analyzing this
observed emission from the rims of SNRs, we can address several fundamental 
issues regarding cosmic-ray acceleration. For example, we can probe the 
particular distribution of high-energy electrons along acceleration sites and 
determine if these electrons are gathered in a diffuse plasma, clumped in 
filaments or have accumulated immediately before the forward shock. We can 
also estimate the maximum energy of cosmic-ray electrons accelerated along 
the shock fronts associated with SNRs. Because cosmic-ray electrons and protons
are expected to be accelerated along the shock in a similar manner 
\citep{Ellison91}, these studies can also help us to improve our understanding of the
acceleration of both leptonic and hadronic cosmic-ray particles. There have   
been several recent claims that TeV emission has been detected from three of the 
SNRs mentioned above, namely SN 1006 \citep{Tanimori98}, G266.2$-$1.2
\citep{Katagiri05} and G347.3$-$0.5 
\citep{Muraishi00, Enomoto02} with the CANGAROO (Collaboration of Australia and
Nippon (Japan) for a GAmma Ray Observatory in the Outback) series of telescopes,
specifically CANGAROO-I and CANGAROO-II \citep{Hara93,Kawachi01}.
We note that similar observations made with 
another TeV observatory, the High Energy Stereoscopic System, also known as
H.E.S.S. \citep{Benbow05}, have also detected TeV emission from SN 1006 
\citep{Acero10}, G266.2$-$1.2 \citep{Aharonian05, Aharonian07a} and G347.3$-$0.5 
\citep{Aharonian04, Aharonian06, Aharonian07b}. Currently, there is no consensus 
about the physical 
origin of this TeV emission: some authors had argued that this emission is 
from cosmic microwave background photons which have been upscattered to TeV 
energies via inverse Compton scattering interactions with ultrarelativistic 
cosmic-ray electrons \citep{Muraishi00,Ellison01,Reimer02,Pannuti03,Lazendic04,Lyutikov04,
Ogasawara07, Plaga08,Katz08}. In contrast, 
other authors have instead claimed that this emission is due to neutral-pion 
decay typically at the site of an interaction between an SNR and a molecular 
cloud \citep{Enomoto02, Fukui03, Uchiyama03, Malkov05, Aharonian06, Moraitis07,
Aharonian07a,Berezhko08,Telezhinsky09,Tanaka08,Fang09,Morlino09}. Several authors 
have assumed 
the former interpretation and included the emission detected at TeV energies 
from these SNRs to model processes related to cosmic-ray acceleration by these 
sources over extremely broad ranges of energies such as radio through TeV 
\citep{Allen01,Reimer02,Uchiyama03,Pannuti03,Lazendic04,Ogasawara07}. In 
\citet{Pannuti03}, we studied cosmic-ray electron acceleration by one of the SNRs with an X-ray 
spectrum dominated by non-thermal emission (G347.3$-$0.5) with a 
spatially-resolved broadband (0.5-30.0 keV) analysis of data from observations 
made with {\it ROSAT}, {\it ASCA}, and the Rossi X-ray Timing Explorer 
({\it RXTE}). In this paper, we investigate the phenomenon of cosmic-ray 
electron acceleration by SNRs through an analysis of X-ray and radio observations of the 
northwestern rim complex of the SNR G266.2$-$1.2 made with {\it Chandra} and the
Australia Telescope Compact Array (ATCA), respectively.
\par
\begin{figure}
\epsscale{1.0}
\plotone{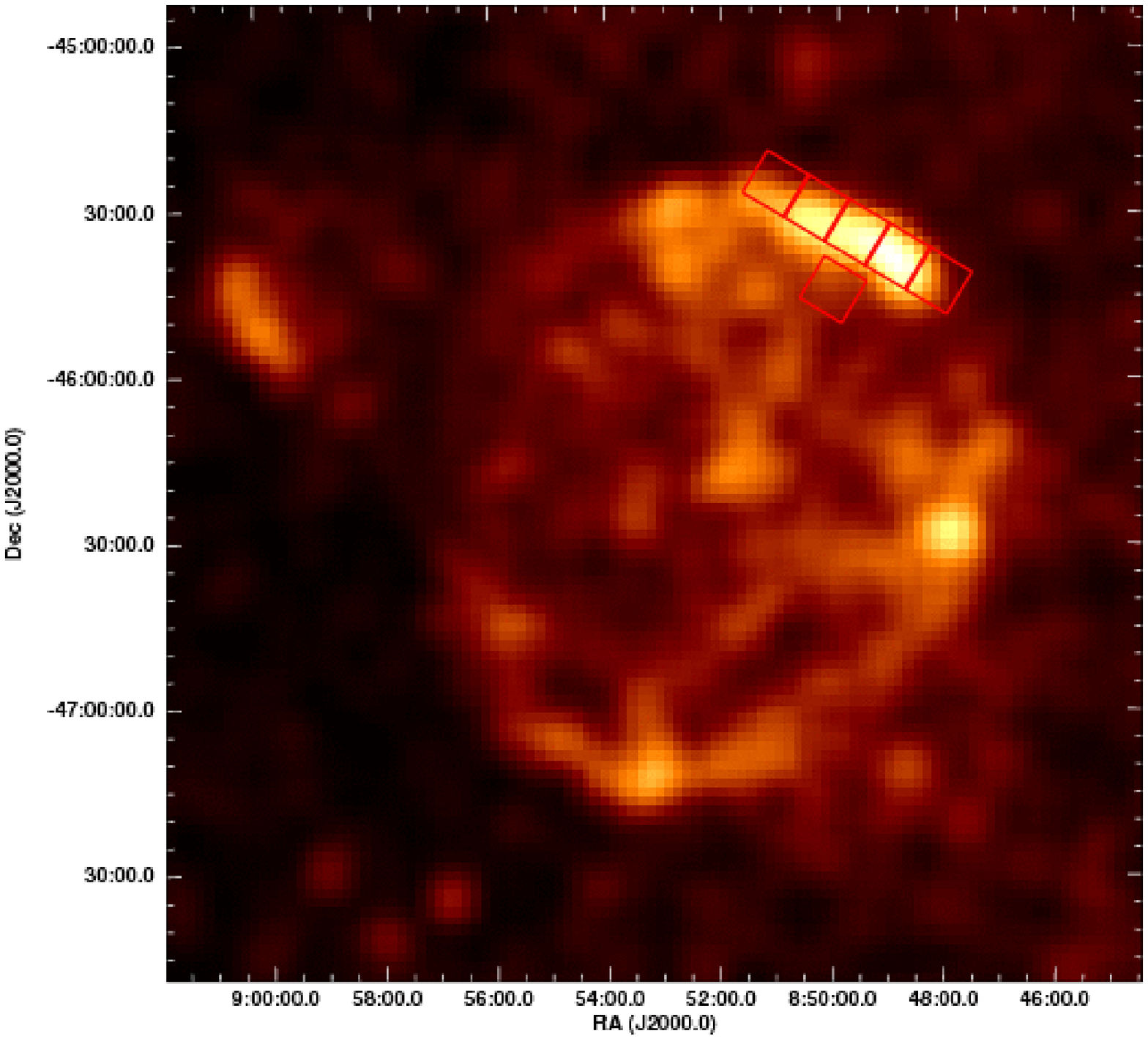}
\caption{A mosaicked X-ray image of the Galactic supernova remnant (SNR) G266.2$-$1.2 
(RX J0852.0-4622) as prepared from all-sky survey images made with the Position Sensitive
Proportional Counter (PSPC) aboard the R\"{o}ntgensatellit (ROSAT). The emission is shown for
photon energies from 1.3 keV to 2.4 keV: we have excluded softer energies to
help reduce the effects of confusing emission from the Vela SNR. The image has
been smoothed with a Gaussian of 1 arcminute and the plotted scale is 0 to 3.2 counts per
pixel (each pixel is approximately 1.7 arcminutes $\times$ 1.7 arcminutes in size). A clear 
shell-like morphology is 
seen: the approximate placement of the Advanced CCD Imaging Spectrometer (ACIS)
chips (namely chips ACIS-S0, -S1, -S2, -S3, -S4 and-I2) for our observation of the northwestern
rim are indicated with red squares. \label{fig1}}
\end{figure}

\citet{Aschenbach98} discovered G266.2$-$1.2 as RX~J0852.0$-$4622, a hard X-ray
source approximately 2$^{\circ}$ in diameter seen in projection against the 
Vela SNR in {\it ROSAT} All-Sky Survey images: this source has also been 
referred to as Vela Z and Vela Jr. in the literature. {\it ROSAT} observations
\citep{Aschenbach98, Aschenbach99} and {\it ASCA} observations
\citep{Tsunemi00,Slane01a,Slane01b} show that the object is shell-like, with 
a luminous northwestern rim and less luminous northeastern, western and southern rims. 
In Figure \ref{fig1} we present a mosaicked map of G266.2-1.2 that was 
generated from observations made with the Position Sensitive Proportional 
Counter (PSPC) aboard {\it ROSAT}. The observed shell-like X-ray morphology
of G266.2$-$1.2 is similar to the X-ray morphologies of other SNRs in
its class, such as SN 1006, G1.9$+$0.3, G330.2$+$1.0 and G347.3$-$0.5. An extreme ultraviolet
(EUV) image made of G266.2$-$1.2 at the wavelength of 83 \AA~shows a shell-like
morphology that broadly matches the observed X-ray morphology \citep{Filipovic01}.
\par
Since its discovery, the age of
G266.2$-$1.2 as well as its distance has been the subject of extensive 
debate in the literature. Much of this debate has centered on determining
whether G266.2$-$1.2 is located at a distance comparable to the distance
to the Vela SNR (estimated to be 250$\pm$30 pc -- \citet{Cha99}) or if
G266.2$-$1.2 lies at a distance beyond the Vela SNR. Distances as low as
$\sim$200 pc and ages as low as $\sim$680-1000 years have been suggested
based on several different observational results. These results include
the following: a high temperature (over 2.5 keV) derived from fits to
the extracted {\it ROSAT} PSPC spectra (suggestive of a high shock velocity)
coupled with the large angular size of the SNR \citep{Aschenbach98};
a claimed detection by the Imaging Compton Telescope (COMPTEL) aboard
the {\it Compton} Gamma-Ray Observatory (CGRO) of $\gamma$-ray line emission 
at 1.157 MeV from the titanium-44 decay chain \citep{Iyudin98}; analyses of 
extracted X-ray spectra of the northwestern rim of G266.2$-$1.2 (as
observed by {\it ASCA} and {\it XMM-Newton}) which suggested an excess
of calcium \citep{Tsunemi00} and emission lines produced by scandium
and titanium \citep{Iyudin05} which are expected from the titanium-44 decay
chain; a putative association between a peak in measured nitrate
abundances in Antarctic ice cores with historical supernovae \citep{Rood79,Motizuki09} 
and the scenario of a very
recent supernova event producing G266.2$-$1.2 \citep{Burgess00}; finally,
age estimates based on the measured widths of the non-thermal X-ray filaments
in the northwestern rim of G266.2$-$1.2 as observed by {\it Chandra} and 
the application of models of synchrotron losses for cosmic-ray electrons
accelerated along this rim \citep{Bamba05a}. However, other studies and 
accompanying analyses have produced arguments that indicate that G266.2$-$1.2 
is significantly more distant (and thus older): for example, fits to
the extracted X-ray spectra of the bright rims of G266.2$-$1.2 as
observed by {\it ASCA} and {\it XMM-Newton} revealed column densities that
are elevated compared to column densities measured toward the Vela SNR
\citep{Slane01a,Slane01b,Iyudin05}. In fact, \citet{Slane01a} and
\citet{Slane01b} argued that G266.2$-$1.2 is physically associated with
a concentration of molecular clouds known as the Vela Molecular Ridge
\citep{May88} which lie at a distance of $\sim$800-2400 pc \citep{Murphy85}.
Other studies have cast doubt on the statistical significance of the 
claimed detections of X-ray and $\gamma$-ray emission lines in extracted
spectra from the northwestern rim of G266.2$-$1.2 
\citep{Schoenfelder00,Slane01a,Hiraga09} as well as associations in general between 
historical supernovae and peaks in nitrate abundances in Antarctic ice
cores \citep{Green04}. Recently, \citet{Katsuda08} estimated an expansion
rate of G266.2$-$1.2 based on two observations made with {\it XMM-Newton}
of the northwestern rim of G266.2$-$1.2 that were separated by 6.5 years.
Those authors derived a low expansion rate of 0.023$\pm$0.006\% for 
G266.2$-$1.2: from this rate they estimated an age of 1.7-4.3 $\times$ 10$^3$
years for the SNR and a distance of 750 pc. We will adopt this distance
to G266.2$-$1.2 for the remainder of this paper. 
\par
In contrast to the extensive debate about the age of G266.2$-$1.2 and its true 
distance, there has been general agreement about the type of supernova that 
produced this SNR, namely a Type Ib/Ic/II event with a massive stellar 
progenitor. This conclusion is based on estimates of the current SNR shell
expansion speed \citep{Chen99} as well as analyses of a central X-ray source 
in this SNR using data from multiple observatories, including {\it ROSAT} 
\citep{Aschenbach98}, {\it ASCA} \citep{Slane01a}, {\it BeppoSAX} 
\citep{Mereghetti01} and {\it Chandra} \citep{Pavlov01}. Based on these 
analyses, it is believed that the central X-ray source CXOU J085201.4$-$461753 
is a radio-quiet neutron star which was formed when the massive stellar 
progenitor of that SNR collapsed. \citet{Pellizzoni02} claimed to have detected 
an H$\alpha$ nebula that may be physically associated with this central source.
However, \citet{Redman05} have argued that the 65-millisecond radio pulsar 
PSR J0855$-$4644 -- seen in projection toward the southeast edge of 
G266.2$-$1.2 -- is instead the neutron star physically associated with this 
SNR. We note that \citet{Redman00} and \citet{Redman02} claimed to have found 
optical emission associated with the outer edge of G266.2$-$1.2, arguing that 
the filamentary nebula RCW 37 -- also known as NGC 2736 and the Pencil Nebula 
-- and the X-ray structure labeled ``Knot D/D'" by \citet{Aschenbach95} are 
associated with G266.2$-$1.2. \citet{Sankrit03} dispute this result, arguing 
that both RCW 37 and D/D' are instead part of the Vela SNR. 
\par
In this paper, we present a detailed spatially-resolved spectroscopic X-ray
study of fine structure in the luminous northwestern rim complex of 
G266.2$-$1.2 as revealed by observations made with {\it Chandra}. This
rim is known to be resolved into both a leading rim and a trailing rim
\citep{Bamba05a, Iyudin05}. These {\it Chandra} observations have been
previously analyzed by \citet{Bamba05a}: in the present paper, we extend
this work in two ways. Firstly, we have measured the widths of the rim structures 
at more locations. Secondly, we also incorporate radio
observations of the northwestern rim complex to help constrain synchrotron
models used to fit X-ray spectra extracted for seven regions within the
complex. The organization of this paper is as follows: in Section 
\ref{ObsSection}, we describe the observations (and the corresponding data
reduction) made of the northwestern rim complex of G266.2$-$1.2 with 
{\it Chandra} (Section \ref{ChandraSubSection}) and the ATCA (Section 
\ref{ATCASubSection}). The data analysis 
and the results are presented in Section \ref{DataResultsSection} where we 
describe the observed fine X-ray emitting structures in the rim complex and 
the measured widths of the observed structures (Section 
\ref{RimStructureSubSection}) as well as fits to the extracted spectra of several
regions in the rim complex using different models for non-thermal X-ray 
emission (Section \ref{NonThermalSubSection}). In Section \ref{MaximumSection} we
present an estimate for the maximum energy of the cosmic-ray electrons 
accelerated along the northwestern rim complex of G266.2$-$1.2: we compare 
this estimate to estimates published for the maximum energies of cosmic-ray 
electrons accelerated by other SNRs. Lastly, we present our conclusions in 
Section \ref{ConclusionsSection}.

\section{Observations and Data Reduction\label{ObsSection}}

\subsection{{\it Chandra} Observations and Data 
Reduction\label{ChandraSubSection}}

The northwestern rim complex of G266.2$-$1.2 was observed by {\it Chandra}
between 2003 January 5-7 in two separate epochs (ObsIDs 3846 and 4414):  
details of these observations are provided in Table \ref{Table1}. 
X-ray emission from the rim complex was imaged with the Advanced CCD Imaging 
Spectrometer (ACIS) camera aboard {\it Chandra} \citep{Garmire03} in Very Faint 
Mode with a focal plane temperature of $-$120$^{\circ}$ Celsius. The ACIS is composed 
of a 2 $\times$ 2 (ACIS-I) and a 1 $\times$ 6 (ACIS-S) array of CCDs. Six of 
these ten detectors (ACIS-I2, -S0, -S1, -S2, -S3 and -S4) were live
during the observations. Each 1024 pixel $\times$ 1024 pixel CCD has a
field of view of 8.'4 $\times$ 8.'4. The angular resolution of the 
{\it Chandra} mirrors and ACIS varies over the observed portion of the
northwestern rim of G266.2$-$1.2 from approximately $0.''5$ at the
aimpoint to $7''$ for a region that is located approximately $10'$ off 
axis. The maximum on-axis effective area for the mirrors and the ACIS-S3 
CCD is approximately 670~cm$^2$ at 1.5 keV: for energies between
0.4 and 7.3 keV, the on-axis effective area is greater than 10\% of this
value. The fractional energy resolution (FWHM/E) between these energies
ranges from about 0.4 to 0.03 respectively. The sensitive energy bands
and energy resolutions of the other five CCDs used in the two
observations are typically worse than those of the ACIS-S3 CCD. In Figure 1 
we have plotted the positions of the fields of view of 
each active ACIS CCD during this observation of the northwestern rim complex
over the complete {\it ROSAT} image of the SNR for the energy range of 1.3 through 
2.4 keV.  
\par
The ACIS data for both observations were reduced using the {\it Chandra}
Interactive Analysis of Observations (CIAO\footnote{See http://cxc.harvard.edu/ciao/.}) 
\citep{Fruscione06} Version 4.0.1 (CALDB Version 3.4.3). The steps taken in producing
new EVT2 files for both observations are outlined as follows: first, the CIAO tool \texttt{destreak}
was used to remove anomalous streak events recorded by the ACIS-S4 chip. Next,
the CIAO tool \texttt{acis\_process\_events} was used to generate a new EVT2 file
(with the parameter settings of \texttt{check\_vf\_pha=yes} and \texttt{trail=0.027}
implemented as recommended for processing observations made in Very Faint mode). This
EVT2 file was then filtered based on grade (where grades=0,2,3,4,6 were selected),
``clean" status column (that is, all bits set to zero) and photon energy (all photons
with energies between 0.3 and 10 keV were selected). Good time intervals supplied
by the pipeline were also applied: in addition, a light curve for the observation
was prepared and examined for the presence of high background flares during the 
observation using the CIAO tool \texttt{ChIPS}. No variations were seen  at the level
of 3$\sigma$ or greater: we therefore conclude that background flares do not affect
the dataset at a significant level. For the final EVT2 files generated for ObsIDs
3846 and 4414, the effective exposure times weere 39363 seconds and 33917 seconds,
respectively, for a total effective exposure time of 73280 seconds.
The two EVT2 files were combined to produce a single co-added image with an
elevated signal-to-noise ratio using the CIAO tool \texttt{dmmerge}. 
We also extracted spectra for seven regions of interest
in the rim complex for both of the observations. Specifically, we considered two regions 
which were sampled by the ACIS-S2 CCD,
four regions which were sampled by the ACIS-S3 CCD and one region which was sampled
by the ACIS-S4 CCD. For each region, we prepared a pulse-height amplitude (PHA)
file, an ancillary response file (ARF) and a response matrix file (RMF) using the
contributed CIAO script {\it specextract}: this tool automatically accounts for 
the build-up of absorbing material on the instruments and adjusts the ARFs accordingly. 
The spectra were grouped to a
minimum of 15 counts per bin to improve the signal-to-noise ratio. We performed 
simultaneous joint fits on the extracted spectra for each region from
both images: analyses of the spectra of these regions was conducted using the 
X-ray spectral analysis software package {\it XSPEC} Version 12.4.0ad \citep{Arnaud96}. 
The results of
these spectral analyses are presented in Section \ref{NonThermalSubSection}. 

\subsection{ATCA Observations and Data Reduction\label{ATCASubSection}}

The entire angular extent of G266.2$-$1.2 was observed with the ATCA on 
14 November 1999 in array configuration EW214 at the wavelengths of 20 and 13~cm 
($\nu$=1384  and 2496~MHz, respectively). Details of these observations are presented and
described by \citet{Stupar05}: to summarize, the observations were done in so-called ``mosaic''
mode  which consisted of 110 pointings over a 12 hour period. The sources 1934$-$638 and
0823$-$500 were used for primary and secondary calibration, respectively.  
\par
The \textsc{miriad} \citep{Sault06} and \textsc{karma} \citep{Gooch06} software packages 
were used for reduction and analysis. Baselines formed with the sixth ATCA antenna were
excluded,  as the other five antennas were arranged in a compact configuration. Both images
feature a resolution 
of 120\arcsec\ and an estimated root-mean-square noise of 0.5~mJy/beam. Also, both images are 
heavily influenced by sidelobes originating from the nearby strong radio source CTB 31 
(RCW 38). Nevertheless, these images allow us to more efficiently study the larger scale
components of this SNR.
\par
In the present paper, we concentrate on the radio properties of the northwestern
rim of G266.2$-$1.2: the portion of the rim complex that was sampled by the {\it Chandra}
observation 
most closely corresponds to the region denoted as ``N2" by \citet{Stupar05}. For a 
region along this rim with an angular size of 75 square arcminutes, we measure integrated
flux densities at the frequencies of 1420 MHz and 2400 MHz of $S$$_{\rm{1420~MHz}}$ =
0.410$\pm$0.062 Jy and $S$$_{\rm{2400~MHz}}$ = 0.305$\pm$0.046 Jy, respectively.
Complementary observations made of this region at two other frequencies, namely 843 MHz
and 4800 MHz \citep{Stupar05}, yield additional flux density estimates of 
$S$$_{\rm{843~MHz}}$ = 0.651$\pm$0.098 Jy and $S$$_{\rm{4800~MHz}}$ = 
0.215$\pm$0.032 Jy,
respectively. Based on these flux density values, we estimate a spectral index\footnote{Defined
such that $S$$_{\nu}$ $\propto$ $\nu$$^{-\alpha}$.} $\alpha$ = 0.62$\pm$0.21.
We also interpolate an integrated flux density at 1 GHz for this rim complex of
$S$$_{\rm{1 GHz}}$ = 0.544 Jy and an average surface brightness of 
$\bar{\Sigma}$$_{\rm{1~GHz}}$ =  
7.25$\times$10$^{-3}$ Jy/arcmin$^2$. We will use these estimates for the spectral index and the
average surface brightness when fitting the extracted X-ray spectra using synchrotron models as
described in Section \ref{NonThermalSubSection}.

\section{Data Analysis and Results\label{DataResultsSection}}

\subsection{Fine Structure of the Northwestern Rim Complex and Widths of
Observed Filaments in the Leading and Trailing 
Rims\label{RimStructureSubSection}}

\begin{figure}
\plotone{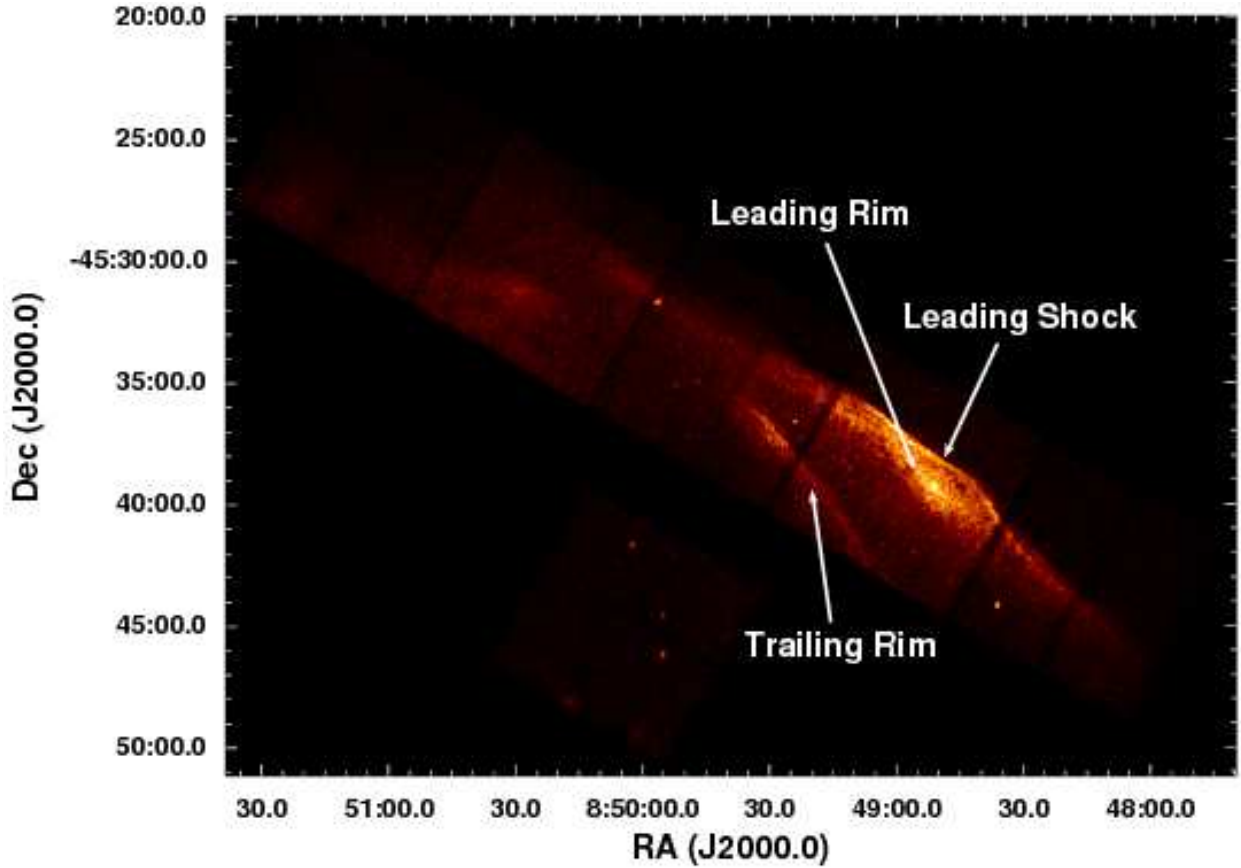}
\caption{An ACIS image of the northwestern rim complex of G266.2-1.2: the 
emission is shown for the energy range 1.0 through 5.0 keV. To  
boost the signal-to-noise, we have combined datasets from both observations
to produce a merged image: we have also smoothed the image 
with a Gaussian of 1.5 arcseconds. Bright filamentary structure, a leading and 
trailing rim and a bright leading shock are all apparent in this image and are labeled.\label{fig2}}
\end{figure}

\begin{figure}
\plotone{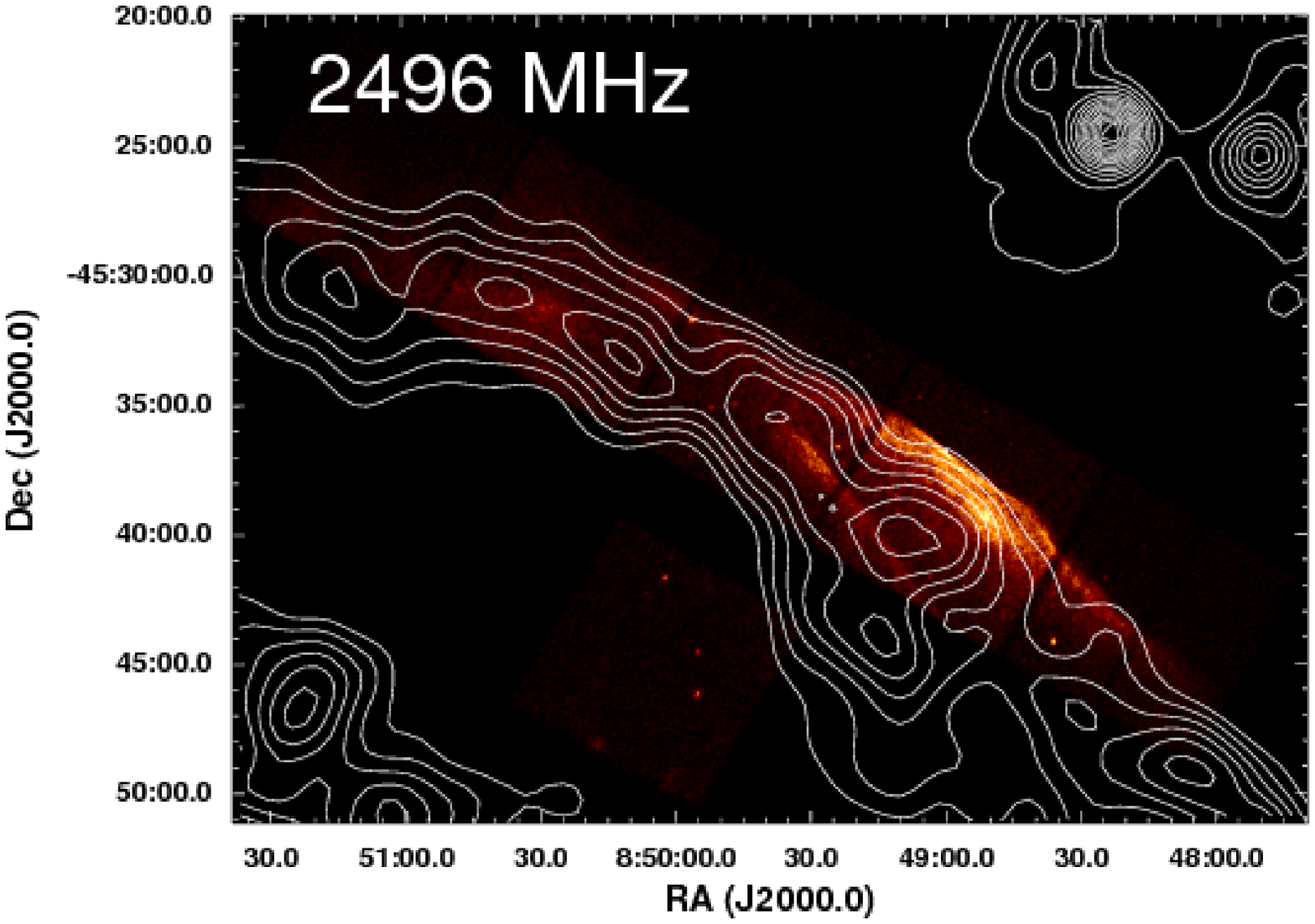}
\caption{An ACIS image of the northwestern rim complex of G266.2$-$1.2 for the
same energy range shown in Figure \ref{fig2} with radio contours 
(depicting emission observed at  2496 MHz) overlaid. The contour levels range from 
0.005 to 0.070 Jy/beam in steps of 0.005 Jy/beam. The resolution of the radio observation
is 120$\arcsec$. See Section \ref{ObsSection}.\label{fig3}}
\end{figure}

\begin{figure}
\plotone{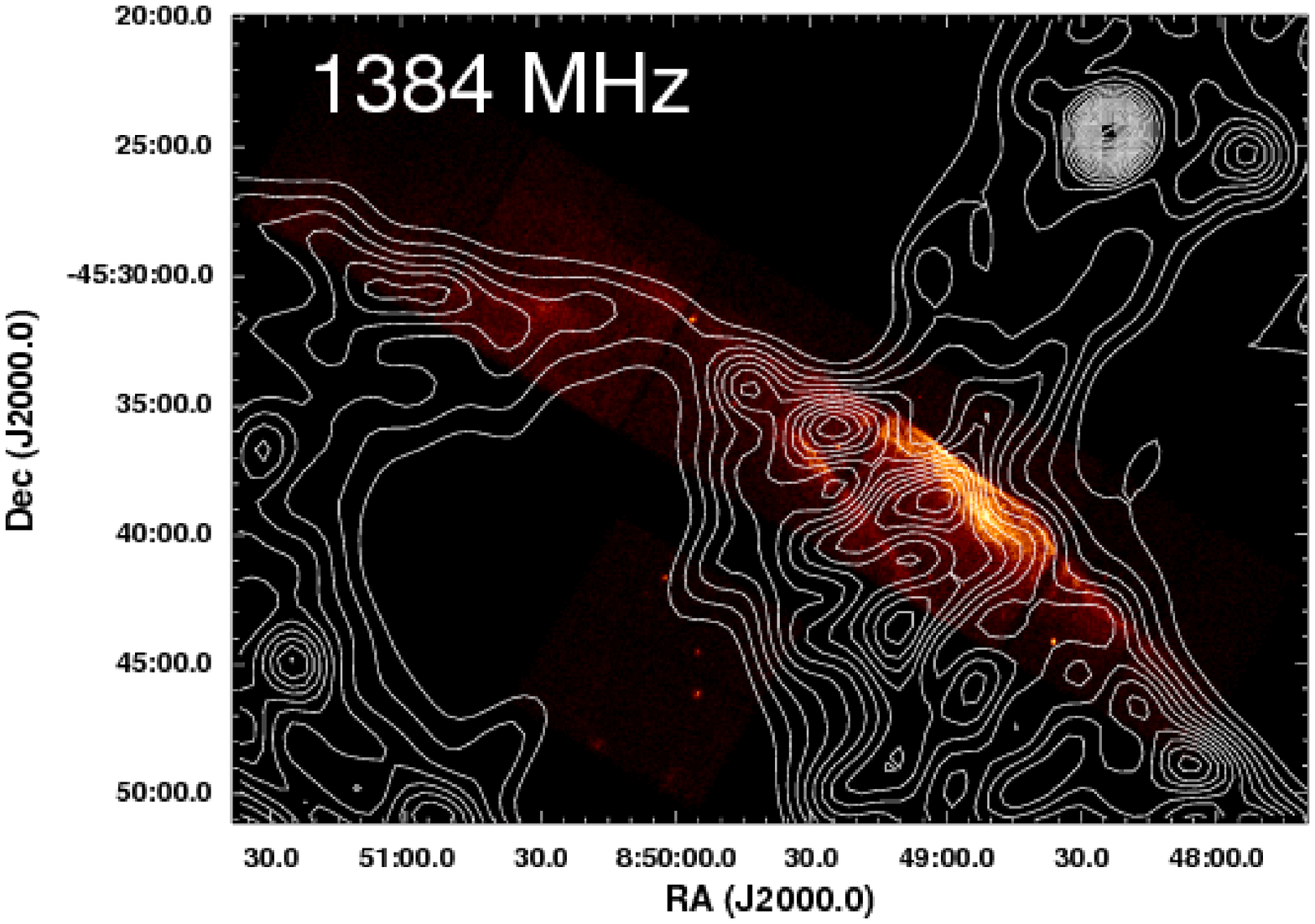}
\caption{An ACIS mage of the northwestern rim complex of G266.2$-$1.2 for the same 
energy range shown in Figure \ref{fig2} with radio contours 
(depicting emission observed at 1384 MHz) overlaid. The contour levels range from 
0.005 to 0.105 Jy/beam in steps of 0.005 Jy/beam. The resolution of the radio observation
is 120$\arcsec$. See Section \ref{ObsSection}.\label{fig4}}
\end{figure}

In Figure \ref{fig2}, we present our reduced, co-added and exposure-corrected {\it Chandra} image
of the northwestern rim complex of G266.2$-$1.2 for the energy range from 1 to 5 keV.
The image has also been smoothed with a Gaussian of $1.5''$.
We have labeled several salient features of this rim complex, including a thin but prominent
leading shock, a leading rim and a trailing rim: these features were noticed by prior
studies of this rim complex \citep{Bamba05a, Iyudin05} and the high angular resolution 
capabilities of {\it Chandra} are essential for a detailed spatially-resolved spectroscopic
study of this complex. The X-ray emission from this rim complex is known to be 
rather hard. In Figures \ref{fig3} and \ref{fig4} we present the same image of the X-ray emission
with the radio contours overlaid at the frequencies of 2496 MHz and 1384 MHz,
respectively. 
\par
As noted in Section \ref{IntroductionSection}, filaments with hard X-ray spectra have been
observed along the leading rims of several Galactic SNRs, including  SN 1006 
\citep{Long03,Bamba03}, Cas A \citep{Gotthelf01,Vink03,Bamba05b}, Kepler 
\citep{Bamba05b,Reynolds07}, Tycho \citep{Hwang02,Bamba05b}, RCW86 \citep{Bamba05b}, 
G266.2$-$1.2 \citep{Bamba05a}, G330.2$+$1.0 \citep{Park09} and 
G347.3$-$0.5 \citep{Uchiyama03, Lazendic04}.  
It is generally accepted that these filaments originate from non-thermal (synchrotron)
emission from cosmic-ray electrons accelerated along the SNR shock and
that the widths of these filaments can help constrain estimates of both
the downstream magnetic field and the maximum energy of the synchrotron-emitting
electrons \citep{Bamba04,VinkJ04, Vink05}. In Table \ref{Table3} we present
published estimates of widths for the observed filaments for five of the 
SNRs mentioned above: we have also adopted distance estimates to 
calculate the corresponding linear widths of the filaments as well. 
Remarkably, the ranges of values for the filament widths are rather
consistent with each other (though it should be noted that estimates of
distances to Galactic SNRs remain significantly uncertain): similar findings 
were reported by \citet{Bamba04} and \citet{Bamba05b}. This result suggests that a narrow range 
of widths of the filaments of SNRs (particularly the young SNRs considered 
here) is dictated by the dynamics of SNR evolution. We note that in contradiction 
the filament widths seem to be much larger for G347.3$-$0.5 by nearly an order of
magnitude for an assumed distance of 6.3 kpc \citep{Lazendic04}, but the
contradiction disappears for the much closer distance of 1 kpc claimed by
other authors \citep{Fukui03,CassamChenai04}. 
\par
\begin{figure}
\rotate
\epsscale{0.8}
\plotone{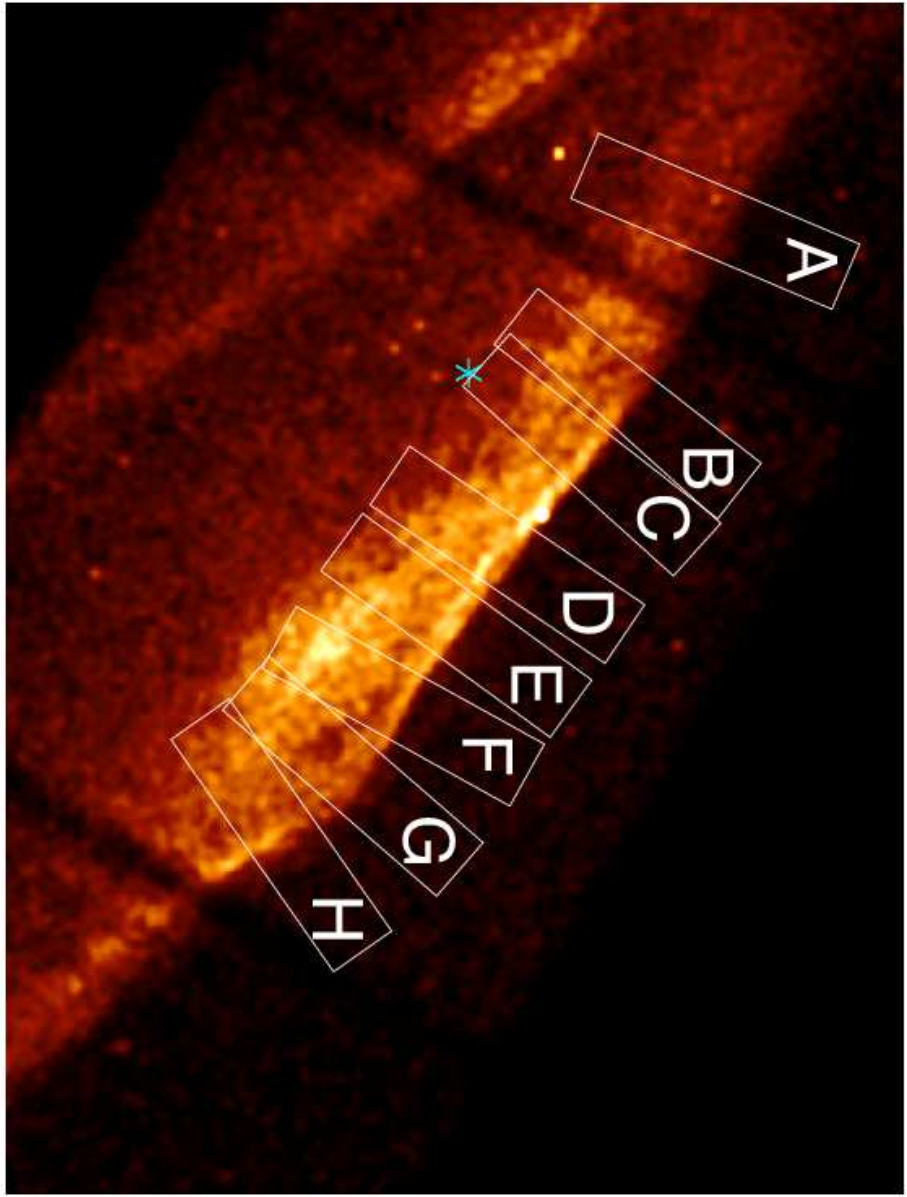}
\caption{An ACIS image of the northwestern rim of G266.2$-$1.2 for the same energy range
shown in Figure \ref{fig2}. 
The bright filaments along the outer edge of the remnant
are evident.  The eight rectangles labeled A--H are the eight regions used
to measure the widths of the bright X-ray-emitting filaments.  The boxes
are centered lengthwise on the portion of the rim with the largest surface
brightness and are oriented at the angles that minimized the upstream
filament widths $l_{1}$. The asterisk marks the location of the optical axis
of the telescope. See Section \ref{RimStructureSubSection} and 
Table~\ref{Table4}).\label{figimg}}
\end{figure}
\begin{figure}
\plotone{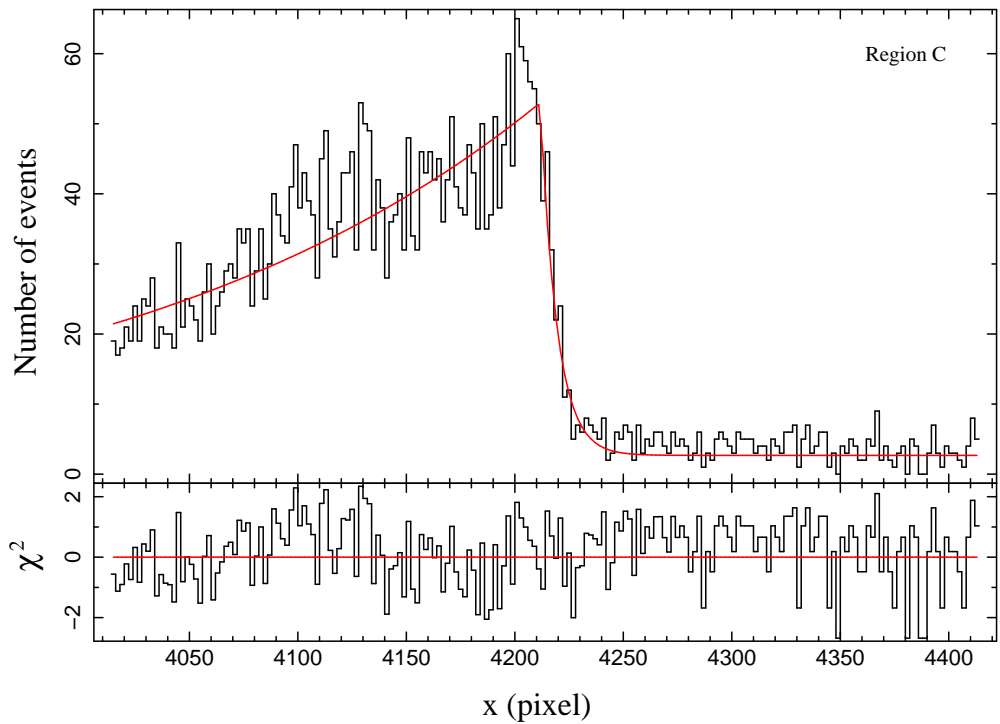}
\caption{A histogram of the number of events in Region C as a function
of the position $x$ along the long axis of this region (top panel). The position 
angle of this axis is $312\arcdeg$.  The best-fit profile (see
Equation~\ref{eqnprofile}) is also shown in the top panel.  The bottom panel
shows the difference between the data and the model divided by the
1~$\sigma$ uncertainties in the data.\label{fighist}}
\end{figure}

To compare the widths of the observed rims seen in the northwestern
rim complex of G266.2$-$1.2 by our {\it Chandra} observation with the
widths of the filaments seen in the other SNRs, we performed the following
fitting procedure. We note that this work extends the analysis conducted by 
\citet{Bamba05a} of filaments associated with the rim complex by analyzing
a total of eight regions (compared with three presented in that work) and
regions located on the ACIS-S2 and -S3 chips (instead of just the ACIS-S3
chip). 
At several locations along the bright northwestern rim (see
Figure~\ref{figimg}), we characterized the length scales over which the X-ray
surface brightness increases from the nominal level outside the SNR to
the peak value along the rim. At each location, a $49\arcsec \times 197\arcsec$
(100~pixel $\times$ 400~pixel) rectangular extraction box was defined. This
particular box size was chosen to strike a balance between choosing a size
small enough to avoid artifically broadening the width because the shock front has
some curvature (that is, it is not entirely linear) but large enough to 
contain enough photons for detailed spatial fitting. The rotation angles of
the boxes were adjusted until the long axes of the boxes appeared to be
perpendicular to the filament: next, the Declination coordinates of the
centers of the boxes were adjusted until these centers were within one pixel
of the locations of the peak surface brightness. Once these criteria were met,
the iterative fitting process began: the events in each box were
extracted and a one-dimensional histogram along the long axis of the box was
created (e.g.\ see Fig.~\ref{fighist}).  Each histogram was fit with the
function
\begin{equation}
  y = 
    \left\{
      \begin{array}{ll}
        A \exp \left( - \frac{x - x_{0}}{l_{1}} \right) + C
          & {\rm if}\ x \ge x_{0} \\
        A \exp \left( + \frac{x - x_{0}}{l_{2}} \right) + C
          & {\rm if}\ x < x_{0} \\
      \end{array}
    \right.
    \label{eqnprofile}
\end{equation}
where $y(x)$ is the total number of events at $x$,
$x$ is the coordinate along the long direction of the box
($x$ is larger at larger radii), $x_{0}$ is the location of the peak surface
brightness, $l_{1}$ is
the length scale over which the surface brightness increases from the
nominal level ``outside'' the remnant to the peak value at the rim, $l_{2}$
is the length scale over which the surface brightness decreases ``inside''
the SNR from the peak value at the rim, $A$ is the peak number of
events and finally $C$ is the nominal number of background events. 
The five free parameters in the fits include $x_{0}$, $l_{1}$, $l_{2}$, $A$ and $C$.
\par
\begin{figure}
\plotone{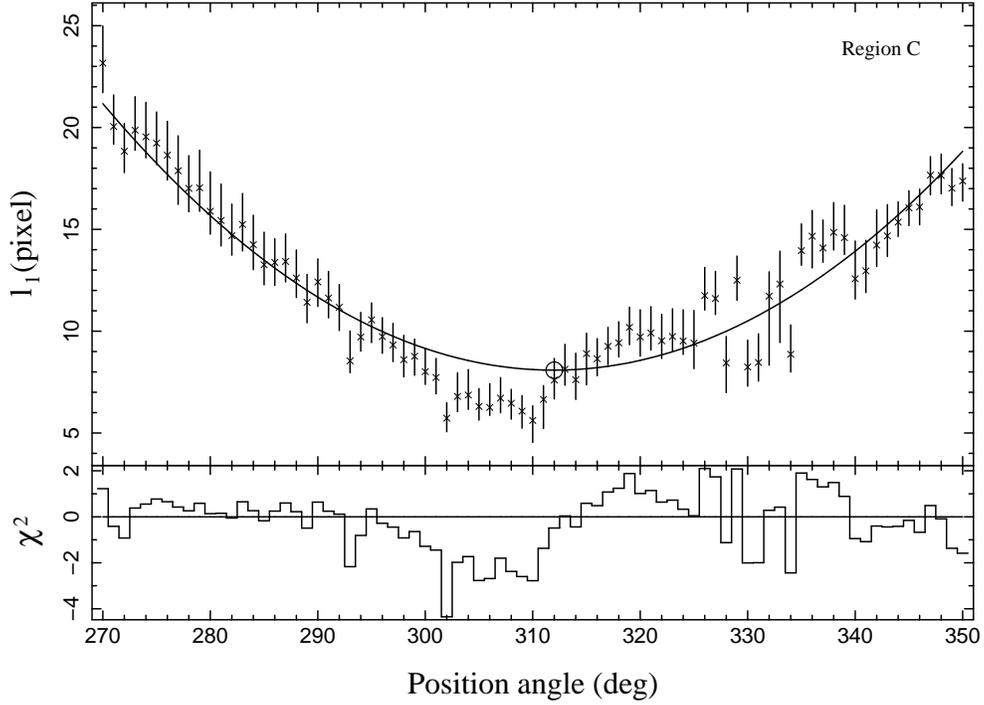}
\caption{
The best-fit upstream width $l_{1}$ for Region C as a function of the position angle
of the long axis of the extraction region for this region.  The smooth curve in the top panel
is the parabola (see eqn~\ref{eqnparabolla}) that best fits the data.  The
lower panel shows the differences between the data and the model divided by
the 1~$\sigma$ uncertainties in the data.  The circle marks the best-fit position
angle of $312^{\circ} \pm 19^{\circ}$.\label{figangle}}
\end{figure}
\begin{figure}
\plotone{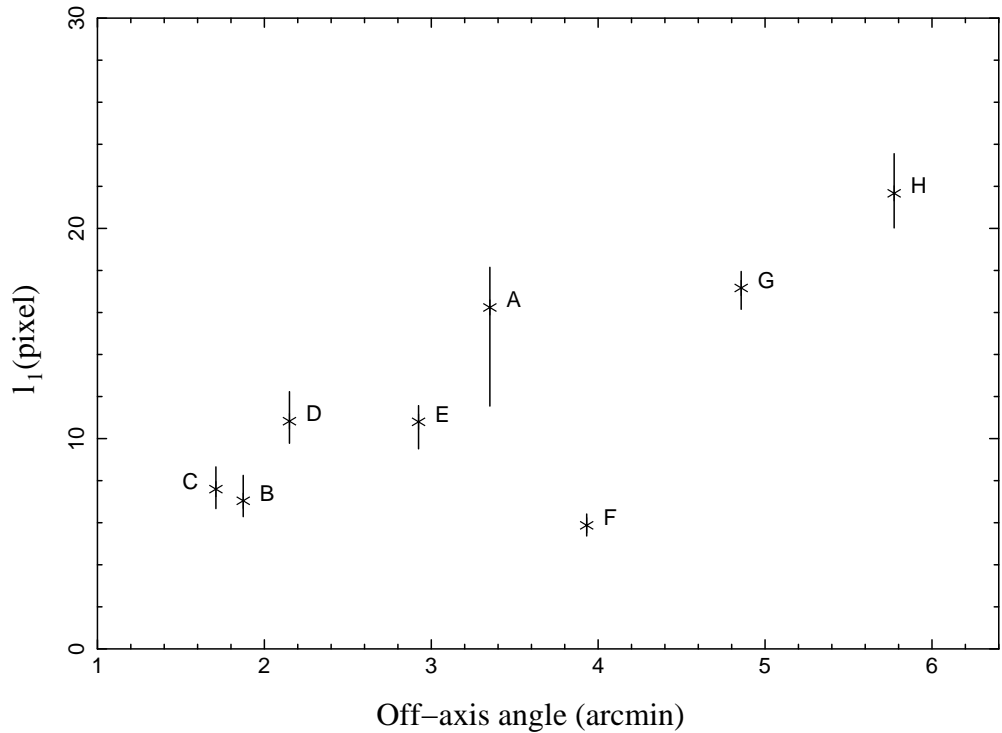}
\caption{
The correlation between the best-fit upstream filament widths $l_{1}$ and the angular
separations of the centers of the extraction regions and the optical axis of
the telescope.  A correlation is expected since the point-spread function of
the telescope increases as a source is moved away from the optical axis.
\label{figoffaxis}}
\end{figure}

To ensure that the length scales $l_{1}$ and $l_{2}$ are not artificially
increased by using boxes whose long sides are not perpendicular to the rim,
several rotation angles were used.  The rotation angles $\theta$ were varied
from $\theta_{0} - 40^{\circ}$ to $\theta_{0} + 40^{\circ}$ in one degree
steps. The angles $\theta_{0}$ represent the best initial guesses at the
rotation angles of the boxes.  At each rotation angle, the data were
re-extracted, histogramed and fitted as described above.  At each location
on the rim, the best-fit rotation angle $\theta_{1}$ was obtained by fitting
the results of $l_{1}$ versus $\theta$ with a quadratic function of the form
\begin{equation}
  l = a_{0} + a_{1} \theta + a_{2} \theta^{2}
  \label{eqnparabolla}
\end{equation} 
(e.g.\ see Figure~\ref{figangle}).  The angles $\theta_{1}$ at which the 
quadratic functions have minima and the best-fit values for $l_{1}$ at
these angles are listed in Table~\ref{Table4}.  The boxes are oriented
in Figure~\ref{figimg} at the best-fit angles $\theta_{1}$. 
As shown in Figure~\ref{figoffaxis}, there is a correlation between the
lengths $l_{1}$ and the off-axis angles $\phi$, which correspond to 
the off-axis angles of the center of the extraction box.  This correlation
demonstrates that the point-spread function of the {\sl Chandra} mirrors
contributes significantly to the fitted length scales at least for the
locations that are far from the optical axis.  The length scales
$l_{1}$ of Regions B and C may be relatively
free of the effects of the mirrors.  In this case, the actual length scale
of the filaments may be about 7~pixels (about $3.5''$).
\par
The best-fit values for $l_{2}$ and $\theta_{2}$, which are obtained in
the same way as the values for $l_{1}$ and $\theta_{1}$, are not listed in
Table~\ref{Table4} because the length scales $l_{2}$ are not necessarily 
the length scales over which the X-ray surface brightness decreases inside the 
leading shock. It is very likely that the geometry of the SNR may not be as simple
as a thin spherical shell, which would make estimates of $l_{2}$ less 
physically meaningful. Some
of the emission that appears to be ``inside'' the SNR in
Figure~\ref{figimg} may be part of the same irregular surface that produces
the leading shock.  This emission may appear to be inside the remnant because
it is seen in projection. Any emission from an irregular surface that is
seen in projection can (and did) result in artificially large fitted values for the
length scales $l_{2}$.  The length scales $l_{1}$ should be relatively
immune to such effects.
\par 
In Table \ref{Table5} we list our derived length scales $l_{1}$ (in both
arcseconds and parsecs) for the eight analyzed filaments and compare these
with estimates for length scales of the filaments associated with five historical
SNRs that were measured and presented by \citet{Bamba03} and \citet{Bamba05b}.
Lastly, we note the recent work
of \citet{Pohl05} who claim that the widths of these filaments are tied to the rapid
decline in the strength of the magnetic field downstream from the turbulence rather than the
energy losses of the radiating electrons. 

\subsection{Modeling Non-Thermal X-ray Emission from Fine 
Structure in the Northwestern Rim Complex\label{NonThermalSubSection}}

\begin{figure}
\plotone{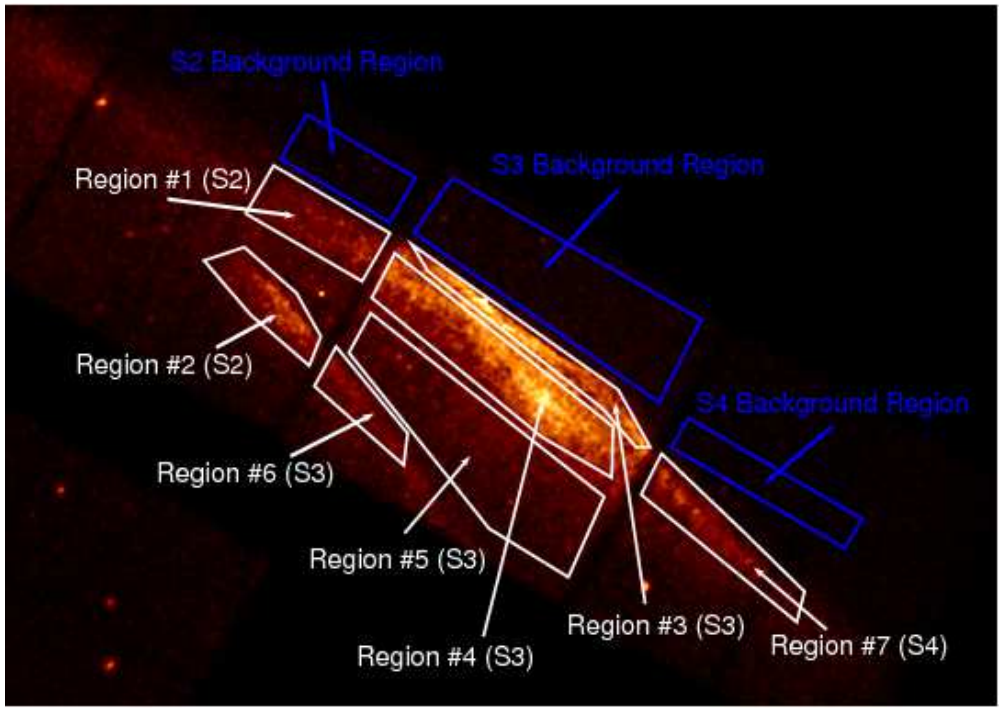}
\caption{The extraction regions for the spectral analysis of the nonthermal X-ray
emission from the northwestern rim complex of G266.2$-$1.2 (see Section 
\ref{NonThermalSubSection}). We have indicated both the seven regions of
spectral extraction (in white) as well as the regions where background spectra were
extracted (in blue). Also see Table \ref{Table6}.\label{G266regionsfig}}
\end{figure}

As described in Section \ref{ObsSection}, we have extracted spectra from seven
regions in the rim complex and in Table \ref{Table6} we list relevant properties of these
regions. With the {\it a priori} knowledge that the X-ray emission from the rim 
complex is predominately non-thermal, we attempted to fit the extracted spectra
using three different non-thermal models, described below.
To create background spectra for the purposes of our spectral modeling, we extracted
spectra from regions located just ahead of the leading edge of the rim complex on 
each chip. This choice of background helps us to account for diffuse 
thermal emission
seen from the Vela SNR in projection against the rim complex itself 
\citep{Lu00,Aschenbach02}. In all cases, we
have used the photoelectric absorption model PHABS \citep{Arnaud96} to account for the 
absorption along
the line of sight to the northwestern rim complex. The regions considered for
spectral analysis (both source regions and the background regions) are depicted
in Figure \ref{G266regionsfig}. We note that this work extends previous spectral analysis
of this rim complex that was presented by \citet{Bamba05a} by considering the properties
of regions sampled by the ACIS-S2, -S3 and -S4 chips instead of just the ACIS-S3 chip.
\par
The first model we considered was a simple power-law. This model has traditionally
been used to fit the X-ray spectra extracted from {\it ROSAT} and {\it ASCA} observations of the 
X-ray luminous rims of Galactic SNRs that feature non-thermal emission, such as SN 1006
\citep{Koyama95} and G347.3$-$0.5 \citep{Slane99}, as well as G266.2$-$1.2
\citep{Slane01a,Slane01b}. The typical values for the photon indices derived from fits with 
power-law models to the spectra of these bright rims are $\Gamma$ $\sim$ 2.1--2.6. In the case of 
G266.2$-$1.2, \citet{Slane01a} derived a photon index of 2.6$\pm$0.2 when fitting the spectrum of 
the northwestern rim, and photon indices of $\Gamma$=2.6$\pm$0.2 and 
$\Gamma$=2.5$\pm$0.2 for fits to the extracted spectra of the northeastern rim 
and the western rim, respectively. In their spectral analysis of dataset obtained
from the {\it XMM-Newton} observation of this rim complex, \citet{Iyudin05} derived a photon
index of $\Gamma$ $\sim$ 2.6 for fits to the higher energy emission ($E$ $\geq$ 0.8 
keV) emission from the entire complex: this power-law component was combined with thermal
components (to model emission separately from the Vela SNR and from G266.2$-$1.2) for
modeling the entire broadband emission (0.2 $\leq$ $E$ $\leq$ 10.0 keV). The results of our 
fits to the spectra of the seven regions in the rim complex using this model are given in 
Table \ref{Table7}: for each region we give the best fit values for the column density $N$$_H$,
photon index $\Gamma$, the normalizations of each fit and for each ObsID (along with 90\%
confidence ranges for each of these parameters), the ratio of the values of the $\chi$$^2$ to the
number of  degrees of freedom (i.e., the reduced $\chi$$^2$ value) and the absorbed 
and unabsorbed fluxes. We have performed our fits over the energy range of 1.0 through 5.0 
keV for each extracted spectra.
\par
\begin{figure}
\epsscale{0.80}
\plotone{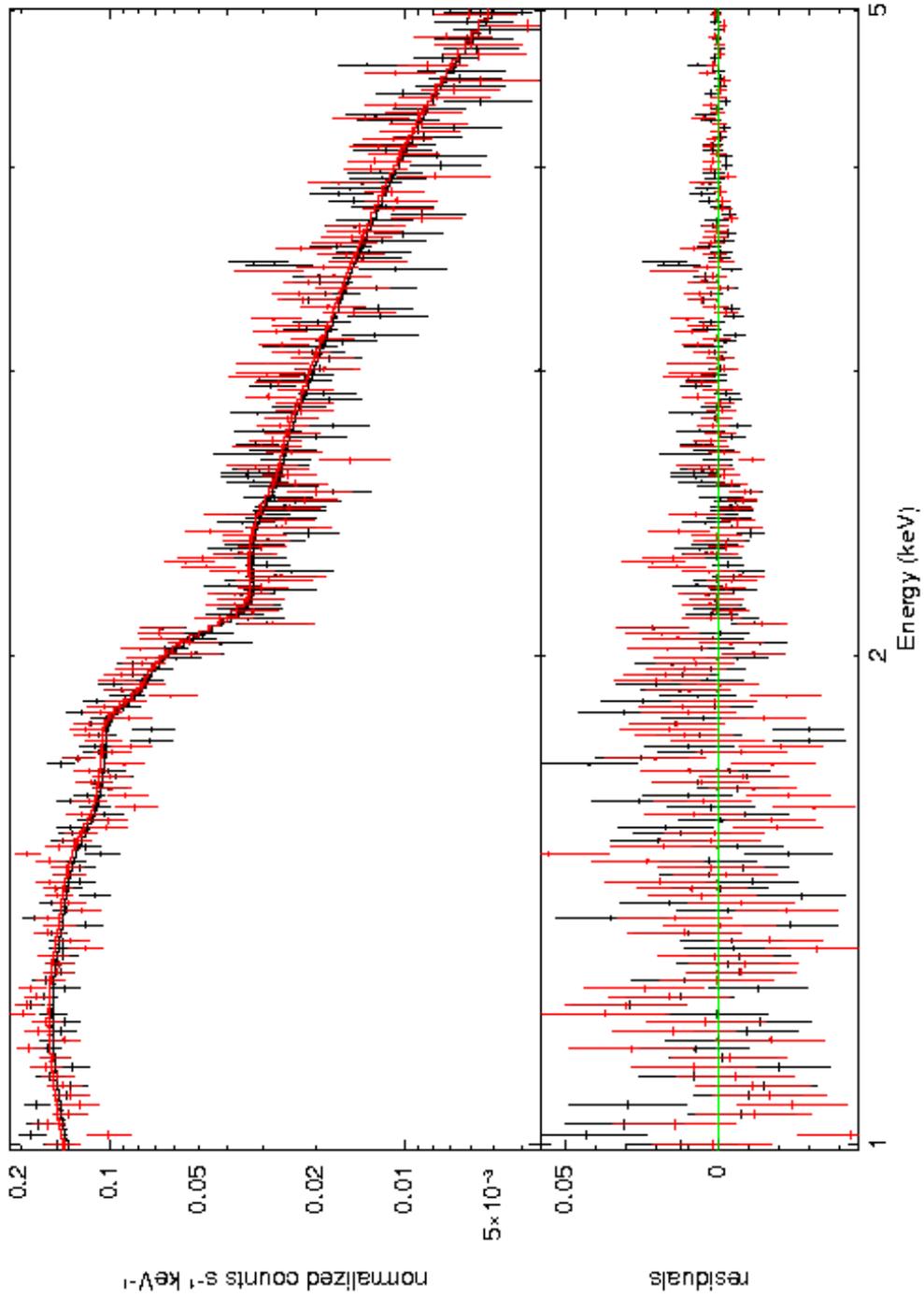}
\caption{Extracted spectra for Region \#3 as fit with a PHABS*Power
Law Model (see Table \ref{Table7}). The spectrum from ObsID 3816 is
shown in black while the spectrum from ObsID 4414 is shown in red.
\label{ShockSpectrafig}}
\end{figure}

\begin{figure}
\plotone{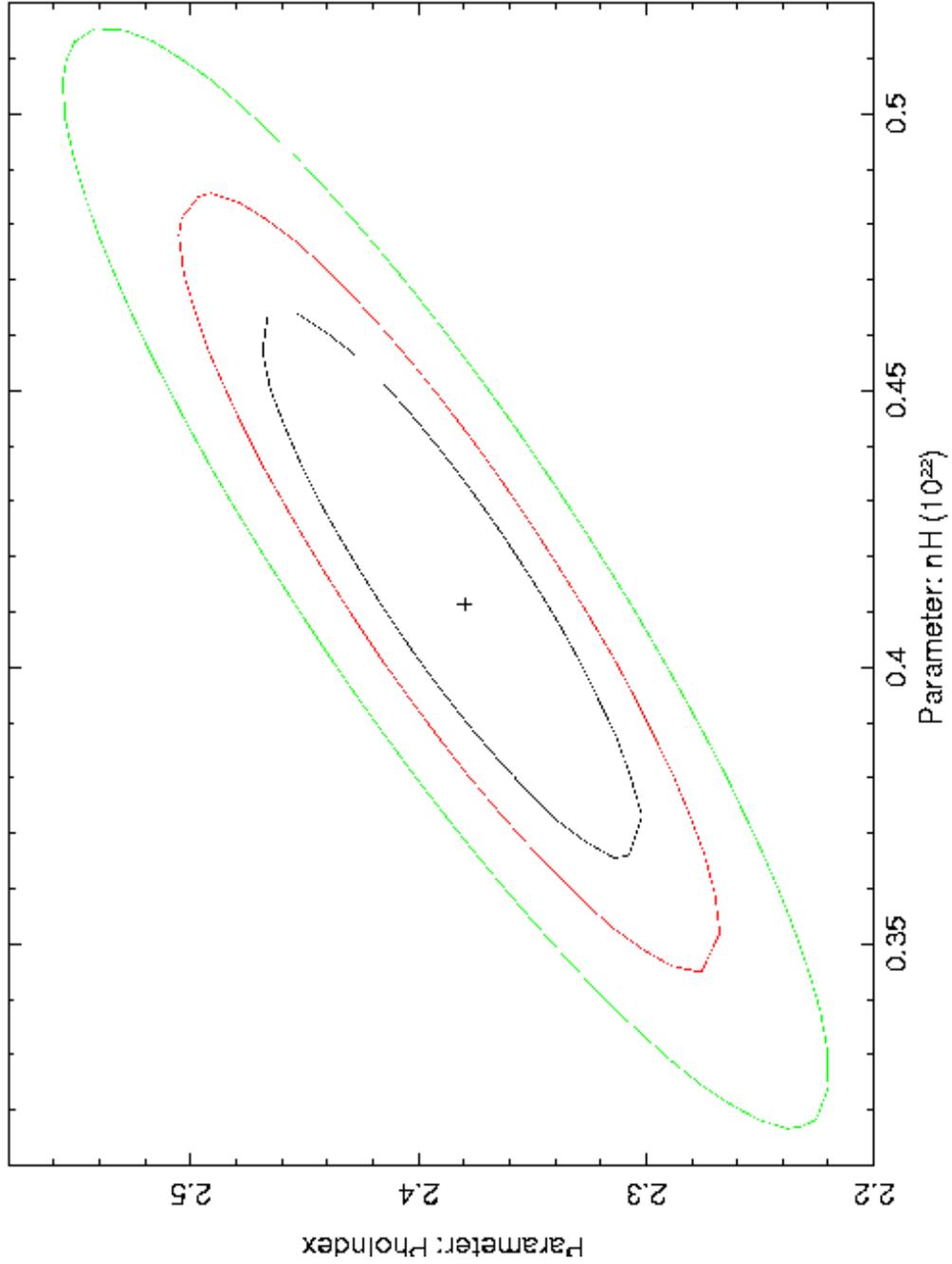}
\caption{Confidence contours (at the 1$\sigma$, 2$\sigma$ and 3$\sigma$ levels) 
for the PHABS*Power Law fit to the extracted spectra
for Region \#3 (see Figure \ref{ShockSpectrafig} and Table \ref{Table7}).
\label{ConfConfig}}
\end{figure}

 
Statistically acceptable fits have been found for all regions using the power-law model. Our
derived photon indices range from approximately 2.38 through 2.70 (in broad agreement with the
photon index value measured for the whole complex by \citet{Slane01a}) and differences are seen
in the values of the photon index for different regions. Several regions in the rim complex may
actually be physically associated based on similar values for the derived photon indices: one pair
of plausibly associated regions are Regions \#3 and \#7, both
of which feature indices of $\Gamma$ $\sim$ 2.38 (the lowest values of photon indices for any
of the regions in the rim complex) and appear to form a clearly-defined leading shock. 
Similarly,  Regions \#1 and \#4 appear to form a feature, seen in projection which trails
behind the leading shock (though Region \#1 is a leading structure 
in the portion of the rim complex sampled by the ACIS-S2 chip).  
Finally, Regions \#2 and \#6 form a trailing rim that is even more clearly distinct
from the leading shock; we suspect that this rim also appears to lie interior to 
the leading shock simply because of projection effects. In their
analysis of {\it XMM-Newton} observations of this SNR, \citet{Iyudin05} first mentioned the
presence of this trailing rim though the superior angular resolution of {\it Chandra} more
clearly resolves this feature as distinct from the leading shock, as noted by \citet{Bamba05a}. 
We interpret the slight range of values of the photon index to possibly indicate the presence of 
variations in the shock conditions of accelerated
electrons within the complex: the lower index of the leading rim indicates that
cosmic-ray electrons are being accelerated to higher energies than at the sites of the
regions with higher indices. We will continue to discuss variations in the shock conditions in
different portions of the rim complex when we describe our spectral fits with the synchrotron
models below.
For illustrative purposes, in Figures \ref{ShockSpectrafig} and \ref{ConfConfig} we present the
extracted spectra of one region (namely Region \#3) as fit with the PHABS$\times$Power Law
model and  confidence contours for this fit, respectively.   
\par
The other two models that we used to fit the extracted spectra, SRCUT and
SRESC, assume a synchrotron origin for the hard X-ray
emission observed from the SNR. Both of these models have been described in detail
elsewhere but we provide brief descriptions of each one here. The SRCUT model 
\citep{Reynolds98, Reynolds99, Hendrick01} describes a synchrotron spectrum from an 
exponentially cut-off power-law distribution of electrons in a uniform magnetic field. This 
model assumes an electron energy spectrum $N$$_e$($E$) of the form
\begin{equation}
N_e (E) = K E^{-\Gamma} e^{\frac{-E}{E_{\rm{cutoff}}}}
\label{SyncEqn}
\end{equation}
where $K$ is the normalization constant derived from the observed flux
density of the region of the SNR at 1 GHz, $\Gamma$ is defined as 2$\alpha$+1 
(where $\alpha$ is the radio spectral index) and $E$$_{\rm{cutoff}}$ is the
maximum energy of the accelerated cosmic-ray electrons. The SRESC model
\citep{Reynolds96, Reynolds98} describes a synchrotron spectrum from an
electron distribution limited by particle escape above a particular energy. This model
describes shock-accelerated electrons in a Sedov blast wave which are interacting
with a medium with uniform density and a uniform magnetic field. The SRESC model
takes into account variations in electron acceleration efficiency with shock obliquity 
as well as post-shock radiative and adiabatic losses. \citet{Dyer01} and
\citet{Dyer04} have described successfully applying the SRESC model to fit extracted
X-ray spectra from the bright rims and the interior of SN 1006 using data from
observations made by both {\it ASCA} and {\it RXTE}. A crucial feature of both the 
SRESC model and the SRCUT model is that a resulting fit made with either of these models  
can be compared with two observable properties of an SNR, namely its flux density at 1 
GHz and radio spectral index $\alpha$. We note that one of the fit parameters of
these two models is the cutoff frequency $\nu$$_{\rm{cutoff}}$ of the synchrotron
spectrum of the accelerated cosmic-ray electrons. This frequency is defined as 
the frequency at which the flux has dropped from a straight power law by a factor 
of ten for
SRCUT and by a factor of six for SRESC. We can express $\nu$$_{\rm{cutoff}}$ as
(see, for example, \citet{Lazendic04})
\begin{equation}
\nu_{\rm{cutoff}} \approx 1.6 \times 10^{16} \left(
\frac{B_{\rm{\mu G}}}{10~\mbox{$\mu$G}} \right) \left( 
\frac{E_{\rm{cutoff}}}{\mbox{10 TeV}} \right)^2 \mbox{Hz},
\label{Eeqn}
\end{equation}
where $B$$_{\rm{\mu G}}$ is the magnetic field strength of the SNR in $\mu$G
and it is assumed that the electrons are moving perpendicular to the magnetic field.
Based on the value for $\nu$$_{\rm{cutoff}}$ returned by the SRESC and SRCUT models
as well as its normalization $K$, we can model the synchrotron spectrum of the
shock-accelerated cosmic-ray electrons associated with SNRs and estimate the
maximum energy $E$$_{\rm{cutoff}}$ of the shock-accelerated electrons. We emphasize
here that to make this estimate of the cutoff energy, either the magnetic field
needs to be measured or a specific value for the magnetic field needs to be assumed.
\par
In applying the SRESC and SRCUT models, the observed radio properties of the
SNR -- namely the flux density at 1 GHz and $\alpha$ -- are needed to constrain the X-ray
spectral fits in a meaningful manner. Like many of the other  
characteristics of G266.2$-$1.2 that were described in Section \ref{IntroductionSection}, 
the radio properties of this SNR have proven to be controversial as well. To investigate the radio
properties of this SNR, \citet{Combi99} used data from the deep continuum survey of the 
Galactic plane made with the Parkes 64 meter telescope at a frequency of 2.4 GHz and with an 
angular resolution of 10.4 arcmin as conducted and described by \citet{Duncan95}. 
\citet{Combi99} combined this dataset with additional 1.42 GHz data obtained with the 30 meter
telescope of the
Instituto Argentino de Radioastronomia at Villa Elisa, Argentina: the half-power beamwidth
of this instrument at 1.42 GHz is approximately 34 arcminutes. Based on these observations, 
\citet{Combi99} claimed to have detected extended limb-brightened features associated with this
SNR at the frequencies of 2.4 and 1.42 GHz and argued that the radio
morphology of this SNR closely corresponds to the X-ray morphology. Different
conclusions were reached by \citet{Duncan00}, however, who performed their own analysis of 
the 2.4 GHz radio data and also considered data from new radio observations
made at 1.40 GHz and 4.85 GHz with the Parkes telescope: the latter data was obtained
as part of the Parkes-MIT-NRAO (PMN) survey \citep{Griffith93} and the angular resolutions  
of these observations were 14.9 and 5 arcminutes, respectively. \citet{Duncan00}
argued that the extended features described by
\citet{Combi99} are more likely to be associated with the Vela SNR than with G266.2$-$1.2 itself. 
By considering the radio emission detected within a circle corresponding to the 
X-ray extent of G266.2$-$1.2, \citet{Duncan00} measured a spectral index 
$\alpha$ = 0.40$\pm$0.15 (where $S$$_{\nu}$ $\propto$ $\nu$$^{-\alpha}$) for the
northern half of the SNR using the method of ``T-T" plots as described by \citet{Turtle62}
and argued that this spectral index value was applicable to all of G266.2$-$1.2.
They also estimated the integrated fluxes from this SNR at the frequencies of
2.42 and 1.40 GHz to be 33$\pm$6 and 40$\pm$10 Jy, respectively. 
By extrapolating their measured integrated flux to a frequency of 1 GHz using the above
spectral index value, \citet{Duncan00} estimated the integrated flux from G266.2$-$1.2
at that frequency to be 47$\pm$12 Jy and from this value calculated an average surface
brightness of this SNR at 1 GHz to be $\Sigma$$_{\rm{1 GHz}}$ to be 
6.1$\pm$1.5 $\times$ 10$^{-22}$ W m$^{-2}$ Hz$^{-1}$ sr$^{-1}$ = 
6.1$\pm$1.5 $\times$ 10$^4$ Jy sr$^{-1}$. The broadband radio observations presented by
\citet{Stupar05} of G266.2$-$1.2 more clearly revealed a shell-like
radio morphology for this SNR: in particular, enhanced radio emission at the northwestern
rim complex. Using the average surface brightness calculated for this rim complex at
1 GHz (see Section \ref{ATCASubSection}) and the measured angular sizes of each
region, we calculated the corresponding flux density at 1 GHz for that region. The
calculated flux densities for each region are provided in Table \ref{Table6}. 
We then performed our spectral analysis with the SRESC and SRCUT models using these
calculated flux densities for the normalizations. 
These frozen normalizations are necessary to help make meaningful interpretations of fits to the
X-ray spectra of these features assuming a synchrotron origin to the emission. Certainly,
local variations are expected in the corresponding flux density at 1 GHz for each region:
newer radio observations made with higher angular resolution are required to clearly
identify such variations and measure the corresponding flux densities. 
\par
In Tables \ref{Table8} and \ref{Table9} we present the results of our fits to the extracted spectra 
using the SRCUT and
the SRESC model. For each fit we have listed the fit parameters including column density
$N$$_H$, spectral index $\alpha$, cutoff frequency $\nu$$_{\rm{cutoff}}$ (along with 90\%
confidence intervals for each of these parameters), the normalization value used, the ratio
of $\chi$$^2$ to the number of degrees of freedom for each fit and the corresponding absorbed
and unabsorbed fluxes of each region. 

\section{Estimates of the Maximum Energies of Cosmic-Ray Electrons 
Accelerated by G266.2$-$1.2 and Other SNRs\label{MaximumSection}}

   Based on the the fit values for the cutoff frequencies derived from
fitting the SRCUT model to the spectra of the seven regions, we have used
Equation~\ref{Eeqn} to calculate the maximum energy $E_{\rm cutoff}$ of
cosmic-ray electrons accelerated at each of these regions.  Following
\citet{Reynolds99} and \citet{Hendrick01}, we have assumed
a field strength $B = 10\ \mu$G, which is expected to be a lower limit on
the actual magnetic field strength.  In Table~\ref{Table10} we list our
computed estimates for $E_{\rm cutoff}$ for each region and each assumed
normalization: our estimates for $E_{\rm cutoff}$ for the different regions
range from 30 to 40~TeV with the highest estimated value corresponding to
the portion of the leading shock that corresponds to Region~\#7.  The higher
observed value for $E_{\rm cutoff}$ for this region may indicate that
cosmic-ray acceleration is the most robust at this location and that
cosmic-ray electrons are accelerated to their highest energies in the narrow
regions along the shock front.  Once again we point out that the {\it
Chandra} observations have revealed a possible range of spatial variations
in the acceleration of cosmic-ray electrons along the entire rim complex.
\par
In Tables~\ref{Table11} and \ref{Table12} we present listings of published
estimates for the cutoff frequencies of cosmic-ray electrons accelerated by
a sample of SNRs in the Galaxy and the Large Magellanic Cloud (LMC),
respectively: in each case these estimates have been derived using the SRCUT
model.  Most of these estimates have been taken from \citet{Reynolds99}
and \citet{Hendrick01} (Tables~\ref{Table11} and \ref{Table12}, respectively).
We have included in this list several SNRs (like
Kes~73, 3C~396, G11.2$-$0.3 and RCW~103) considered by \citet{Reynolds99} 
which are now known to be associated with central X-ray sources which
contribute confusing hard emission that was not spatially resolved from the
SNR itself by the {\sl ASCA} observations. Therefore, these estimates for
$E_{\rm cutoff}$ must be interpreted with caution.  In addition, we have
augmented the lists from those two papers with published estimates from
other papers which have considered other SNRs. Using Equation~\ref{Eeqn} and
again assuming a magnetic field $B = 10\ \mu$G, we have calculated a value
for $E_{\rm cutoff}$ for each SNR. Lastly, we have added to
Table~\ref{Table11} our own estimate for $\nu_{\rm cutoff}$ and $E_{\rm
cutoff}$ for G266.2$-$1.2 as described in this paper to help generate an
up-to-date listing of values for $E_{\rm cutoff}$ for SNRs both in the
Galaxy and in the LMC.
\par
   All of the estimates presented here for values of $E_{\rm cutoff}$ must
be interpreted with a great deal of caution, given known significant
uncertainties in our understanding about the true shape of the electron
spectrum and about the magnetic field strength.  As shown in
Equation~\ref{SyncEqn}, the model SRCUT is based on the assumption that the
electron spectrum is a power law with an exponential cut off. If the
cosmic-ray energy density at the forward shock is large enough, then the
shock transition region will be broadened \citep{Ellison91, Berezhko99}.
As a result, cosmic-ray spectra do not have power-law distributions: instead,  
the spectra flatten with increasing energy \citep{Bell87,Ellison91,Berezhko99}.
Evidence of such
curvature has been reported for Cas~A \citep{Jones03}, RCW~86 \citep{Vink06} 
and SN~1006 \citep{Allen08b}.  If magnetic fields in
remnants are $\sim 100\ \mu$G or larger, as has been reported for several
young remnants, then the electron spectrum should be steepened due to
radiative losses (e.g., \citet{Ksenofontov10}). 
Furthermore, theoretical analyses presented by other authors \citep{Ellison01,
Uchiyama03,Lazendic04,Zirakashvili07} have used an alternative form for the cutoff (that is,
$\exp(-(E/E_{\rm cutoff}))^{s}$) where the range of values for $s$ have
included 1/4, 1/2, 1 or 2 (though \citet{Zirakashvili07} have
argued in favor of the use of $s = 2$).  Taken altogether, it is clear that
the simple spectral form of the accelerated electrons in Equation~\ref{SyncEqn}
(as featured in such models as SRCUT and SRESC) is likely to be a
significant over-simplification of the true shapes of the spectra. We
comment that the data presently available for G266.2$-$1.2 may not be
sensitive enough for very rigorous models of the shape of the electron
spectrum. Because the synchrotron emissivity function for a monoenergetic
collection of electrons is quite broad \citep{Longair94}, the shape of
the synchrotron spectrum will be a very broadly smeared version of the shape
of the electron spectrum at the energy where the electron spectrum is
cutoff.  The synchrotron spectrum can also be broadened if the electron
spectrum or magnetic field strength vary within a spectral extraction
region. We also comment that the cutoff frequency is only one of the
parameters in the fit that control the shape of the spectrum: our fits also
include a photon index and an absorption column density and the values of
these shape parameters are correlated. If the electron spectrum is curved
and if radiative losses are important at the cutoff energy in the electron
spectrum, then the synchrotron model is clearly over-simplified and should
include additional shape parameters. If the electron spectrum is curved due
to the pressure of cosmic rays or synchrotron losses or if $s < 1$, then the
estimates for the maximum electron energies in Tables~\ref{Table11} and
\ref{Table12} are overestimates. Furthermore, the assumed magnetic field
strength of $10\ \mu$G is clearly an underestimate for some young SNRs,
where the magnetic field is known to be closer to $100\ \mu$G: in such a
case, the estimates for the maximum energies given here are certainly
overestimates and the true maximum energies may be lower still.
\par
   Despite these reasons to expect that the true electron cut-off energies
are lower than the values listed in Tables~\ref{Table11} and \ref{Table12}, it
is still clear from the augmented tables that no known SNR (either Galactic
or in the LMC) appears to be accelerating cosmic-ray electrons to the energy
of the knee feature of the cosmic-ray spectrum (that is, approximately
3000~TeV), just as pointed out by \citet{Reynolds99} in referencing
their Table~11.  From the work in the present paper, we have established
that G266.2$-$1.2 does not appear to be accelerating cosmic-ray electrons to
the knee feature either: we note that the values listed for $\nu_{\rm cutoff}$ 
in Tables~\ref{Table11} and \ref{Table12} are derived from integrated
spectra for these SNRs, while we have considered for G266.2$-$1.2 the region
with the highest value for $\nu_{\rm cutoff}$. Indeed, wide variations in
the values for this quantity have been reported for spatially-resolved
spectral analyses of SNRs like SN~1006 \citep{Rothenflug04,Allen08b} and 
Cas~A \citep{Stage06}.  Also, as stated by 
\citet{Reynolds99}, it must be stressed that these maximum values are most
likely overestimates given the assumption that most or all of the X-ray flux
from an SNR is non-thermal in origin.  Ignoring the contribution of
thermal emission is clearly incorrect for SNRs which are known to feature
emission lines in their X-ray spectra.  We note that additional analyses
published by other authors have provided {\it lower} limits on the maximum
energy of cosmic-ray electrons accelerated by SNRs.  For example, in
analyzing synchrotron emission at infrared wavelengths from Cas~A, \citet{Rho03} 
concluded that the maximum energy of cosmic-ray electrons
accelerated by that SNR was at least 0.2~TeV. Given the significant losses
through synchrotron radiation which are expected from cosmic-ray electrons
as they are accelerated to high energies, it is not very surprising that no
SNR has been observed to be accelerating cosmic-ray electrons to the energy
of the knee feature. The fact that no SNR features an $E_{\rm cutoff}$ value
that is near the knee energy is not unexpectedrefle given that the knee
feature is more clearly associated with cosmic-ray protons and nucleons. Our
analyses place no constraints upon the maximum energy of the nuclei.

\section{Conclusions\label{ConclusionsSection}}

The conclusions of this paper may be summarized as follows:
\par
1) We have conducted an analysis of two X-ray observations made with {\it Chandra} of the
luminous northwestern rim complex of the Galactic SNR G266.2$-$1.2: the combined effective
integration time of the two observations was 73280 seconds. This SNR is a member of the 
class of Galactic SNRs which feature X-ray spectra dominated by non-thermal emission, which is
commonly interpreted to be synchrotron in origin. We have performed an X-ray spatially-resolved
spectral analysis of 
this rim complex to help probe the phenomenon of non-thermal X-ray emission from
SNRs and its relationship to cosmic-ray acceleration by these sources. These observations have
revealed fine structure in the rim complex, including
a sharply defined leading shock and additional filaments. To help constrain the analysis of 
synchrotron emission from this rim complex, we have also considered radio observations made
of G266.2$-$1.2 with ATCA at the frequencies of 1384 MHz and 2496 MHz.
\par
2) We have modeled the length scales of the upstream X-ray emitting
structures located along the northwestern rim complex and sampled by the ACIS-S2, -S3
and -S4 chips. We compared these length scales with published length scales
for other non-thermal X-ray-emitting structures associated with the rims of five historical Galactic 
SNRs. The length scales range from 0.02 to 0.08 pc and are comparable to those measured for 
the older historical SNRs.
\par
3) We have extracted spectra from seven regions that were sampled using the 
ACIS-S array (namely the ACIS-S2, ACIS-S3 and ACIS-S4 chips) and fit the spectra
using several non-thermal models, namely a simple power-law and the synchrotron
models SRCUT and SRESC. Statistically acceptable fits have been derived using 
each model: however, slight variations are seen in derived fit parameters such
as the photon index and cutoff frequency. We interpret the variations of these parameters
to possibly indicate differences in how cosmic-ray electrons are accelerated along 
different sites
of the rim complex. We argue that the lower photon index and higher cutoff frequencies
derived for the two regions that compose the leading shock (our Regions \#3 and \#7) 
are indicative of cosmic-ray electrons being accelerated to the highest energies at these 
locations. 
\par
4) We estimate the maximum energy of cosmic-ray electrons accelerated along this
rim complex to be approximately 40 TeV. This maximum energy is associated with 
Region \#7, a region that forms a portion of the leading shock of the rim complex.
We have also estimated the 
maximum energies of cosmic-ray electrons accelerated by other Galactic SNRs and
LMC SNRs based on published values of cut-off frequencies. As established in prior
works, no known SNR appears to be currently accelerating cosmic-ray electrons
to the knee energy of the observed cosmic-ray spectrum (though this knee feature
is more applicable for cosmic-ray protons and cosmic-ray nucleons). Such a result
is not unexpected, given that significant synchrotron losses from accelerated cosmic-ray 
electrons  are predicted to limit the maximum energies of these accelerated particles
to energies significantly below the energies associated with the knee feature.



\acknowledgments

We thank the referee for many useful comments which helped improve the overall quality of this
paper. T.G.P. acknowledges useful discussions with Steven Reynolds (regarding the
SRCUT model), Kelly Korreck, Jonathan Keohane and Michael Stage. This research has 
made use of NASA's Astrophysics Data System: we have also used data obtained from the
High Energy Astrophysics Science Archive Research Center (HEASARC), which is
provided by NASA's Goddard Space Flight Center. T.G.P. also acknowledges 
support for this work from NASA LTSA grant NAG5-9237.




\clearpage


\clearpage

\clearpage

\clearpage

\clearpage

\clearpage

\clearpage

\clearpage

\clearpage

\clearpage

\clearpage

%






\clearpage

\begin{deluxetable}{lc}
\rotate
\tablecaption{Summary of Prior {\it Chandra} Observations of SNRs with X-ray
Spectra Dominated by Non-Thermal Emission\label{SummaryObsSNRsTable}}
\tablewidth{0pt}
\tablehead{
\colhead{SNR} & \colhead{References}
}
\startdata
SN 1006 & \citet{Long03}, \citet{Bamba03}, \citet{CassamChenai08}, \citet{Allen08b}, 
\citet{Katsuda09a}\\
G1.9$+$0.3 & \citet{Reynolds08}, \citet{Reynolds09} \\
G28.6$+$0.1 & \citet{Ueno03} \\
G266.2$-$1.2 & \citet{Bamba05a}  \\
G330.2$+$1.0 & \citet{Park09} \\
G347.3$-$0.5 & \citet{Uchiyama03, Lazendic04, Uchiyama07} 
\enddata 
\end{deluxetable}

\clearpage

\begin{deluxetable}{lccccccc}
\tablecaption{Summary of {\it Chandra} Observations of the Northwestern
Rim of G266.2$-$1.2\label{Table1}}
\tablewidth{0pt}
\tablehead{
& & & & & & \colhead{Effective}\\
& & & & & & \colhead{Exposure}\\
\colhead{Sequence} & & \colhead{R.A.} & \colhead{Decl.} &
\colhead{Roll} & \colhead{Observation} & \colhead{Time} \\
\colhead{Number} & \colhead{ObsID} & \colhead{(J2000.0)} & 
\colhead{(J2000.0)} & \colhead{(degrees)} & \colhead{Date} 
& \colhead{(seconds)}
}
\startdata
500325 & 3846 & 08 49 09.40 & $-$45 37 42.41 & 30 & 5 January 2003 & 39363 \\
500325 & 4414 & 08 49 09.39 & $-$45 37 42.35 & 30 & 6 January 2003 & 33917
\enddata


\tablecomments{Units of Right Ascension are hours, minutes and seconds,
and units of Declination are degrees, arcminutes and arcseconds.}

\end{deluxetable}

\clearpage
\begin{deluxetable}{lcccccccc}
\rotate
\tabletypesize{\scriptsize}
\tablecaption{Widths of Synchrotron-Emitting X-ray Filaments Associated with 
Galactic SNRs\label{Table3}}
\tablewidth{0pt}
\tablehead{
& & & \colhead{Apparent}\\
& & & \colhead{Angular} & \colhead{Filament} & \colhead{Filament} & \colhead{Filament Width} \\
& \colhead{Distance} & & \colhead{Radius} & \colhead{Width} & \colhead{Width} 
& \colhead{as Fraction of} \\ 
\colhead{SNR} & \colhead{(kpc)} & \colhead{Reference} & 
\colhead{(arcmin)\tablenotemark{a}} &  
\colhead{(arcsec)}  & \colhead{(pc)} & \colhead{SNR Radius} & \colhead{References}
}
\startdata
SN 1006 & 2.2 & (1) & 15 & 4--20 & 0.04--0.20 & 0.0044--0.022 & (2), (3) \\
Cas A & 3.4 & (4) & 2.55 & 3-4 & 0.05 & 0.02-0.03 & (5), (6) \\
Kepler & 4.8 & (7)  & 1.5 & 2-3 & 0.05--0.07 & 0.02--0.03 & (8) \\
Tycho & 2.3 & (9) & 4 & 5 & 0.06 & 0.021 & (8), (10) \\ 
RCW 86 & 2.8 & (11) & 21 & 20 & 0.27 & 0.016 & (8)  \\
G330.2$+$1.0 & 5.0 & (12) & 5.5 & 12--16 & 0.3--0.4 & 0.036--0.048 & (13) \\
G347.3$-$0.5 & 6.3 & (14) & 30 & 20--40 & 0.6--1.2 & 0.011--0.022 & (15), (16) \\
 & 1.0 & (17) & 30 & 20--40 & 0.1--0.2 \\
\enddata
\tablerefs{(1) \citet{Winkler03}, (2) \citet{Long03}, (3) \citet{Bamba03} 
(4) \citet{Reed95}, (5) \citet{Gotthelf01}, (6) \citet{Vink03}, (7) \citet{Reynoso99},
(8) \citet{Bamba05b}, (9) \citet{Hughes00}, (10) \citet{Hwang02}, (11) \citet{Rosado96},
(12) \citet{McClure01}, (13) \citet{Park09}, (14) \citet{Slane99}, 
(15) \citet{Uchiyama03}, (16) \citet{Lazendic04}, (17) \citet{Fukui03}. }
\tablenotetext{a}{From \citet{Green09a,Green09b}.}
\end{deluxetable}

\clearpage

\begin{deluxetable}{cccccc}
  \tablecaption{Widths of X-ray Emitting Features in Northwestern Rim Complex of
  G266.2$-$1.2\label{Table4}}
  \tablewidth{0pt}
  \tablehead{
    \ &
      R.A.\tablenotemark{a} &
      Dec\tablenotemark{a} &
      $l_{1}$\tablenotemark{b} &
      $\theta_{1}$\tablenotemark{b} &
     $\phi$\tablenotemark{c} \\
    Region &
      (J2000.0) &
      (J2000.0) &
      (pixel) &
      (deg) &
      (arcmin)
  }
  \startdata
  A 
  & 08 49 19.44
  & $-45$ 34 51.2                                   
  & $16.24^{+1.91}_{-4.69}$
  & $293 \pm 14$
  & 3.4
  \\
B
  & 08 49 07.37
  & $-45$ 35 52.0
  & $ 7.04^{+1.21}_{-0.75}$
  & $308 \pm 15$
  & 1.9
  \\
C
  & 08 49 03.99
  & $-45$ 36 16.9
  & $ 7.60^{+1.06}_{-0.92}$
  & $312 \pm 19$
  & 1.7
  \\
D
  & 08 48 57.37
  & $-45$ 37 15.6
  & $10.83^{+1.40}_{-1.05}$
  & $304 \pm 9$
  & 2.2
  \\
E
  & 08 48 52.70
  & $-45$ 37 50.9
  & $10.80^{+0.77}_{-1.29}$
  & $306 \pm 21$
  & 2.9
  \\
F
  & 08 48 47.33
  & $-45$ 38 27.8
  & $ 5.88^{+0.53}_{-0.51}$
  & $301 \pm 9$
  & 3.9
  \\
G
  & 08 48 42.69
  & $-45$ 39 02.7
  & $17.17^{+0.78}_{-1.02}$
  & $311 \pm 14$
  & 4.9
  \\
H
  & 08 48 38.77
  & $-45$ 39 52.3
  & $21.67^{+1.88}_{-1.65}$
  & $325 \pm 35$
  & 5.8
  \\
  \enddata

\tablecomments{Units of Right Ascencion are hours, minutes and seconds,
    and units of Declination are degrees, arcminutes and arcseconds.}

\tablenotetext{a}{Coordinates of the center of each extraction box.}
    
\tablenotetext{b}{The uncertainties are at the 1~$\sigma$ confidence level.} 

\tablenotetext{c}{See Section \ref{RimStructureSubSection}.}
  
\end{deluxetable}

\clearpage

\begin{deluxetable}{lcccccc}
\rotate
\tabletypesize{\scriptsize}
\tablecaption{Comparison of Measured Widths of Non-Thermal X-ray Filaments
Associated with Galactic SNRs\label{Table5}}
\tablewidth{0pt}
\tablehead{
& & \colhead{$l$$_1$} & \colhead{$l$$_1$} & \colhead{$l$$_2$} & \colhead{$l$$_2$} \\ 
\colhead{SNR} & \colhead{Filament} & \colhead{($\arcsec$)} & \colhead{(pc)} &
\colhead{($\arcsec$)} & \colhead{(pc)} & \colhead{Reference}
}
\startdata
SN 1006 & \#1 & 1.7 (1.3, 2.2) & 0.02 (0.01, 0.02) & 160 ($>$92) & 1.7 ($>$0.98) & (1) \\
& \#2 & 14 (9.8, 21) & 0.15 (0.10, 0.22) & 93 (73, 130) & 0.99 (0.78, 1.39) \\
& \#3 & 2.3 (0.78, 4.3) & 0.02 (0.01, 0.05) & 150 ($>$56) & 1.60 ($>$0.60)  \\
& \#4 & 0.12 (--) & 0.001 (--) & 150 ($>$78) & 1.60 ($>$0.83) \\
& \#5 & 4.5 (2.2, 8.7) & 0.05 (0.02, 0.09) & 62 (43, 100) & 0.66 (0.46, 1.07) \\
& \#6 & 3.4 (2.4, 5.0) & 0.04 (0.03, 0.05) & 130 (86, 250) & 1.39 (0.92, 2.67) \\
\hline
Cas A & \#1 & $<$0.93 & $<$0.02 & 1.27 (0.96, 1.81) & 0.02 (0.02, 0.03) & (2)  \\
& \#2 & $<$0.80 & $<$0.01 & $<$1.59 & $<$0.03  \\
\hline
Kepler & \#1 & 1.17 (0.87, 1.59) & 0.03 (0.02, 0.04) & 0.93 ($<$1.41) & 0.02 ($<$0.03) & (2) \\
& \#2 & 1.59 (1.17, 2.89) & 0.04 (0.03, 0.05) & 3.09 (3.46, 3.87) & 0.07 (0.08, 0.09) \\
\hline
Tycho & \#1 & 1.18 (1.01, 1.32) & 0.01 (0.01, 0.01) & 5.36 (4.47, 6.12) & 0.06 (0.05, 0.07) & (2) \\
& \#2 & $<$0.80 & $<$0.009 & 1.70 (1.32, 3.15) & 0.02 (0.01, 0.04) \\
& \#3 & $<$0.80 & $<$0.009 & 2.38 (2.20, 2.54) & 0.03 (0.02, 0.03) \\
& \#4 & 0.86 (0.80, 0.93) & 0.01 (0.009, 0.01) & 5.53 (5.00, 6.14) & 0.06 (0.06, 0.07) \\
& \#5 & 1.03 (0.90, 1.35) & 0.01 (0.01, 0.02) & 2.47 (1.93, 3.15) & 0.03 (0.02, 0.04)\\
\hline
RCW 86 & \#1 & 2.39 (1.48, 3.34) & 0.03 (0.02, 0.05) & 20.1 (17.3, 23.8) & 0.27 (0.23, 0.32) & (2) \\
& \#2 & 1.56 (0.49, 4.79) & 0.02 (0.007, 0.07) & 18.2 (11.8, 35.6) & 0.25 (0.16, 0.48) \\
\hline
G266.2$-$1.2 & 1 & 19.0 (12.1, 31.4) & 0.07 (0.04, 0.11) & 31.8 (23.7, 47.1) & 0.12 
(0.09, 0.17) & (3) \\  
             & 2 & 3.68 (2.69, 5.75) & 0.01 (0.01, 0.02) & 65.0 (38.2, 144.6) & 0.24 (0.14, 0.53) \\
             & 3 & \nodata & \nodata & 37.1 (26.5, 60.2) & 0.13 (0.10, 0.22) \\ 
\hline
G266.2$-$1.2 & A & 16.24 (11.53, 18.15) & 0.06 (0.04, 0.07) & \nodata & \nodata & (4) \\
& B & 7.04 (6.29, 8.25) & 0.03 (0.02, 0.03) & \nodata & \nodata \\
& C & 7.60 (6.68, 8.66) & 0.03 (0.02, 0.03) & \nodata & \nodata \\
& D & 10.83 (9.78, 12.23) & 0.04 (0.03, 0.04) & \nodata & \nodata \\
& E & 10.80 (9.51, 11.57) & 0.04 (0.03, 0.04) & \nodata & \nodata \\
& F & 5.88 (5.37, 6.41) & 0.02 (0.02, 0.02) & \nodata & \nodata \\
& G & 17.17 (16.15, 17.95) & 0.06 (0.06, 0.07) & \nodata & \nodata \\
& H & 21.67 (20.02, 23.55) & 0.08 (0.07, 0.09) & \nodata & \nodata \\ 
\enddata
\tablecomments{The linear scales have been calculated using the distances for SN 1006,
Cas A, Kepler, Tycho and RCW 86 listed in Table \ref{Table3}. In the case of G266.2$-$1.2
the linear scales have been calculated using our adopted distance of 750 pc. 
All quoted errors are 90\% confidence values.}
\tablerefs{(1) \citet{Bamba03} (for the 2.0-10.0 keV band), (2) \citet{Bamba05b}
(for the 5.0-10.0 keV band), (3) \citet{Bamba05a} (for the 2.0-10.0 keV band -- the linear scales
have been computed assuming a distance to G266.2$-$1.2 of 750 pc), (4) This paper 
(for the 1.3-5.0 keV band).}
\end{deluxetable}

\clearpage

\begin{deluxetable}{lcccccc}
\tablecaption{Properties of the Regions With Extracted Spectra\label{Table6}}
\tablewidth{0pt}
\tablehead{
& & \colhead{Approximate} & \colhead{Approximate} & \colhead{Approximate} & 
& \colhead{Flux} \\
& & \colhead{Central} & \colhead{Central} & \colhead{Region} & & \colhead{Density}\\
\colhead{Region} & \colhead{ACIS-S} & \colhead{R.A.} & \colhead{Decl.}
& \colhead{Area} & \colhead{Description} & \colhead{at 1 GHz}\\
\colhead{Number} & \colhead{Chip} & \colhead{(J2000.0)} & 
\colhead{(J2000.0)} & \colhead{(arcmin$^2$)} & \colhead{of Location} &
\colhead{(Jy)\tablenotemark{a}}
}
\startdata
\#1 & S2 & 08 49 24 & $-$45 34 52 & 4.95 & Leading Rim & 0.036\\
\#2 & S2 & 08 49 32 & $-$45 36 48 & 4.39 & Trailing Rim & 0.032 \\
\#3 & S3 & 08 48 53 & $-$45 38 05 & 5.15 & Leading Shock & 0.037\\
\#4 & S3 & 08 48 57 & $-$45 38 31 & 10.20 & Leading Rim & 0.074\\ 
\#5 & S3 & 08 49 02 & $-$45 40 34 & 18.10 & Interior Region & 0.131\\
\#6 & S3 & 08 49 17 & $-$45 39 34 & 3.21 & Trailing Rim & 0.023\\
\#7 & S4 & 08 48 24 & $-$45 42 57 & 5.15 & Leading Shock & 0.037\\
\enddata
\tablecomments{Units of Right Ascencion are hours, minutes and seconds,
and units of Declination are degrees, arcminutes and arcseconds.}
\tablenotetext{a}{Estimated from the angular size of each region and the
calculated surface brightness of the northwestern rim complex at 1 GHz. See 
Section \ref{ATCASubSection}.}
\end{deluxetable}

\clearpage

\begin{deluxetable}{lcccccccccc}
\rotate
\tabletypesize{\scriptsize}
\tablecaption{Fits to the Extracted Spectra Using the PHABS$\times$POWER LAW
Model\tablenotemark{a}\label{Table7}}
\tablewidth{0pt}
\tablehead{

& \colhead{Column} & \colhead{Photon} & & & & \colhead{Absorbed} & \colhead{Unabsorbed} 
& \colhead{Absorbed} & \colhead{Unabsorbed}\\
& \colhead{Density} & \colhead{Index} & \colhead{Normalization\tablenotemark{d}} 
& \colhead{Normalization\tablenotemark{d}}
& \colhead{$\chi$$^2$/Degrees} & \colhead{Flux\tablenotemark{e}} & 
\colhead{Flux\tablenotemark{e}} & \colhead{Flux\tablenotemark{e}} 
& \colhead{Flux\tablenotemark{e}} \\
\colhead{Region} & \colhead{$N$$_{\rm{H}}$\tablenotemark{b}} & 
\colhead{$\Gamma$\tablenotemark{c}} & \colhead{(ObsID 3846)} & 
\colhead{(ObsID 4414)} & \colhead{of Freedom} & \colhead{(ObsID 3846)} 
& \colhead{(ObsID 3846)} 
& \colhead{(ObsID 4414)} & \colhead{(ObsID 4414)} 
}
\startdata
\#1 & 0.43$^{+0.09}_{-0.13}$ & 2.68$^{+0.14}_{-0.22}$ & 6.58$^{+1.12}_{-1.38}$ &
6.48$^{+1.12}_{-1.38}$ & 270.20/279=0.97 & 7.63$\times$10$^{-13}$ & 
1.04$\times$10$^{-12}$ & 7.52$\times$10$^{-13}$ & 1.02$\times$10$^{-12}$\\
\#2 & 0.35$^{+0.12}_{-0.10}$ & 2.56$\pm$0.18 & 4.94$^{+1.06}_{-0.86}$ & 
4.94$^{+1.06}_{-0.94}$ & 238.15/263=0.91 & 6.60$\times$10$^{-13}$ & 
8.40$\times$10$^{-13}$ & 6.60$\times$10$^{-13}$ & 8.40$\times$10$^{-13}$ \\
\#3 & 0.41$\pm$0.07 & 2.38$\pm$0.12 & 7.36$^{+1.04}_{-0.96}$ & 7.58$^{+1.02}_{-0.98}$  &
320.67/343=0.93 & 1.09$\times$10$^{-12}$ & 1.42$\times$10$^{-12}$ &
1.12$\times$10$^{-12}$ & 1.43$\times$10$^{-12}$\\
\#4 & 0.41$\pm$0.05 & 2.52$^{+0.09}_{-0.07}$ & 20.20$^{+2.00}_{-1.60}$ & 
20.00$^{+2.00}_{-1.60}$ & 519.06/477=1.09 & 2.67$\times$10$^{-12}$ & 
3.51$\times$10$^{-12}$ & 2.65$\times$10$^{-12}$ & 3.48$\times$10$^{-12}$ \\
\#5 & 0.36$\pm$0.10 & 2.60$^{+0.14}_{-0.22}$ & 7.99$^{+1.61}_{-1.24}$ & 
8.12$^{+1.68}_{-1.37}$ & 406.38/393=1.03 & 1.06$\times$10$^{-12}$ & 
1.36$\times$10$^{-12}$ & 1.08$\times$10$^{-12}$ & 1.38$\times$10$^{-12}$\\
\#6 & 0.39$^{+0.21}_{-0.13}$ & 2.70$^{+0.35}_{-0.22}$ & 2.25$^{+0.75}_{-0.45}$ & 
2.15$^{+0.75}_{-0.39}$ & 203.13/194=1.05 & 2.61$\times$10$^{-13}$ & 
3.47$\times$10$^{-13}$ & 2.50$\times$10$^{-13}$ & 3.32$\times$10$^{-13}$ \\
\#7 & 0.30$\pm$0.14 & 2.38$\pm$0.18 & 5.41$^{+1.69}_{-0.91}$ & 5.22$^{+1.16}_{-0.97}$ & 
310.23/277=1.12 & 8.56$\times$10$^{-13}$ & 1.04$\times$10$^{-12}$ & 8.26$\times$10$^{-13}$
& 1.00$\times$10$^{-12}$   
\enddata
\tablenotetext{a}{Spectra fit over the energy range of 1.0 to 5.0 keV. All 
quoted errors are 90\% confidence values.}
\tablenotetext{b}{In units of 10$^{22}$ cm$^{-2}$.}
\tablenotetext{c}{Defined such that $F$ $\propto$ $E$$^{-\Gamma}$.}
\tablenotetext{d}{In units of 10$^{-4}$ photons keV$^{-1}$ cm$^{-2}$ sec$^{-1}$ at 1 keV.}
\tablenotetext{e}{In units of ergs cm$^{-2}$ sec$^{-1}$.}
\end{deluxetable}

\clearpage
\begin{deluxetable}{lccccccc}
\rotate
\tabletypesize{\scriptsize}
\tablecaption{Fits to the Extracted Spectra Using the PHABS$\times$SRCUT
Model\tablenotemark{a}\label{Table8}}
\tablewidth{0pt}
\tablehead{
& \colhead{Column}  & \colhead{Spectral} & \colhead{Cutoff} \\ 
&  \colhead{Density} & \colhead{Index} & \colhead{Frequency} & & \colhead{$\chi$$^2$/Degrees} & 
\colhead{Absorbed} & \colhead{Unabsorbed} \\
\colhead{Region} & \colhead{$N$$_{\rm{H}}$\tablenotemark{b}} & 
\colhead{$\alpha$\tablenotemark{c}} & \colhead{$\nu$$_{\rm{cutoff}}$\tablenotemark{d}} & 
\colhead{Normalization\tablenotemark{e}} & \colhead{of Freedom} & 
\colhead{Flux\tablenotemark{f}} & \colhead{Flux\tablenotemark{f}}
}
\startdata
\#1 & 0.18$^{+0.31}_{-0.05}$ & 0.54$^{+0.02}_{-0.10}$ & 1.65$^{+1.25}_{-1.25}$ & 0.036 & 
282.55/280=1.01& 7.75$\times$10$^{-13}$ & 8.70$\times$10$^{-13}$ \\
\#2 & 0.17$^{+0.23}_{-0.05}$ & 0.55$^{+0.01}_{-0.08}$ & 1.83$^{+1.27}_{-1.28}$ & 0.032 & 
245.52/264=0.93 & 6.68$\times$10$^{-13}$ & 7.46$\times$10$^{-13}$ \\
\#3 & 0.32$^{+0.09}_{-0.05}$ & 0.52$^{+0.02}_{-0.03}$ & 1.85$\pm$1.55 & 0.037 & 
325.13/344=0.95 & 1.11$\times$10$^{-12}$ & 1.34$\times$10$^{-12}$ \\
\#4 & 0.24$^{+0.20}_{-0.03}$ & 0.52$^{+0.01}_{-0.08}$ & 1.58$\pm$0.98 & 0.074 & 
549.07/478=1.15 & 2.70$\times$10$^{-12}$ & 3.13$\times$10$^{-12}$ \\
\#5 & 0.21$^{+0.19}_{-0.05}$ & 0.59$^{+0.02}_{-0.07}$ & 1.93$^{+1.27}_{-1.28}$ & 0.131 &
410.81/394=1.04 & 1.09$\times$10$^{-12}$ & 1.24$\times$10$^{-12}$  \\
\#6 & 0.17$^{+0.39}_{-0.07}$ & 0.58$^{+0.01}_{-0.15}$ & 1.92$\pm$1.69 & 0.023 & 
212.45/195=1.09 & 2.63$\times$10$^{-13}$ & 2.93$\times$10$^{-13}$ \\
\#7 & 0.21$^{+0.10}_{-0.08}$ & 0.54$^{+0.01}_{-0.03}$ & 2.80$\pm$1.60 & 0.037 & 
314.72/278=1.13 & 8.42$\times$10$^{-13}$ & 9.64$\times$10$^{-13}$ 
\enddata
\tablenotetext{a}{Spectra fit over the energy range of 1.0 to 5.0 keV. All 
quoted errors are 90\% confidence values.}
\tablenotetext{b}{In units of 10$^{22}$ cm$^{-2}$.}
\tablenotetext{c}{Defined such that $S$$_{\nu}$ $\propto$ 
$\nu$$^{-\alpha}$.}
\tablenotetext{d}{In units of 10$^{17}$ Hz.}
\tablenotetext{e}{In units of Jy. See Table \ref{Table6}.} 
\tablenotetext{f}{In units of ergs cm$^{-2}$ sec$^{-1}$. The absorbed and unabsorbed fluxes 
for the
two ObsIDs for each fit were virtually the same: we present one set of values here.}
\end{deluxetable}

\clearpage

\begin{deluxetable}{lccccccc}
\rotate
\tabletypesize{\scriptsize}
\tablecaption{Fits to the Extracted Spectra Using the PHABS$\times$SRESC 
Model\tablenotemark{a}\label{Table9}}
\tablewidth{0pt}
\tablehead{
& \colhead{Column}  & \colhead{Spectral} & \colhead{Cutoff} \\
& \colhead{Density} & \colhead{Index} & \colhead{Frequency} & 
& \colhead{$\chi$$^2$/Degrees} & 
\colhead{Absorbed} & \colhead{Unabsorbed} \\
\colhead{Region} & \colhead{$N$$_{\rm{H}}$\tablenotemark{b}} & 
\colhead{$\alpha$\tablenotemark{c}} & \colhead{$\nu$$_{\rm{cutoff}}$\tablenotemark{d}} & 
\colhead{Normalization\tablenotemark{e}} & \colhead{of Freedom} & 
\colhead{Flux\tablenotemark{f}} & \colhead{Flux\tablenotemark{f}}
}
\startdata
\#1 & 0.39$^{+0.05}_{-0.22}$ & 0.50$^{+0.06}_{-0.02}$ & 2.47$^{+5.33}_{-0.47}$ & 0.036
& 273.06/280=0.98 & 7.51$\times$10$^{-13}$ & 9.90$\times$10$^{-13}$ \\
\#2 & 0.37$^{+0.05}_{-0.25}$ & 0.51$^{+0.07}_{-0.02}$ & 2.59$^{+8.61}_{-0.59}$ & 0.032 
& 245.25/264=0.93 & 6.48$\times$10$^{-13}$ & 8.39$\times$10$^{-13}$ \\
\#3 & 0.51$^{+0.04}_{-0.36}$ & 0.49$^{+0.09}_{-0.01}$ & 2.63$^{+24.67}_{-0.23}$ & 0.037 
& 362.21/344=1.05 & 1.06$\times$10$^{-12}$ & 1.50$\times$10$^{-12}$ \\
\#4 & 0.41$^{+0.04}_{-0.20}$ & 0.48$^{+0.06}_{-0.01}$& 2.86$^{+6.24}_{-0.28}$ & 0.074
& 540.18/478=1.13 & 2.61$\times$10$^{-12}$ & 3.46$\times$10$^{-12}$ \\
\#5 & 0.38$^{+0.06}_{-0.26}$ & 0.56$^{+0.07}_{-0.02}$ & 2.79$^{+10.71}_{-0.69}$ & 0.131
& 413.90/394=1.05 & 1.04$\times$10$^{-12}$ & 1.37$\times$10$^{-12}$ \\
\#6 & 0.39$^{+0.09}_{-0.25}$ & 0.52$^{+0.08}_{-0.03}$ & 1.88$^{+6.12}_{-0.60}$ & 0.023 
& 205.49/195=1.05 & 2.50$\times$10$^{-13}$ & 3.34$\times$10$^{-13}$ \\
\#7 & 0.42$^{+0.03}_{-0.40}$ & 0.50$^{+0.10}_{-0.01}$ & 2.47$^{+31.53}_{-0.07}$ & 0.037 
& 337.09/278=1.21 & 8.10$\times$10$^{-13}$ & 1.07$\times$10$^{-12}$ \\
\enddata
\tablenotetext{a}{Spectra fit over the energy range of 1.0 to 5.0 keV. All 
quoted errors are 90\% confidence values.}
\tablenotetext{b}{In units of 10$^{22}$ cm$^{-2}$.}
\tablenotetext{c}{Defined such that $S$$_{\nu}$ $\propto$ $\nu$$^{-\alpha}$.}
\tablenotetext{d}{In units of 10$^{17}$ Hz.}
\tablenotetext{e}{In units of Jy. See Table \ref{Table6}.} 
\tablenotetext{f}{In units of ergs cm$^{-2}$ sec$^{-1}$. The absorbed and unabsorbed fluxes 
for the
two ObsIDs for each fit were virtually the same: we present one set of values here.}
\end{deluxetable}

\clearpage

\begin{deluxetable}{lcc}
\tablecaption{Estimates from the SRCUT Model for Cutoff Energies $E$$_{\rm{cutoff}}$ of 
Cosmic-Ray Electrons Accelerated at Different Regions of the Northwestern Rim Complex of 
G266.2$-$1.2\label{Table10}}
\tablewidth{0pt}
\tablehead{
& \colhead{Cutoff} & \colhead{Cutoff} \\
&\colhead{Frequency}  & \colhead{Energy} \\
& \colhead{$\nu$$_{\rm{cutoff}}$} & \colhead{$E$$_{\rm{cutoff}}$}\\
\colhead{Region} & \colhead{(10$^{17}$ Hz)} & \colhead{(TeV)} 
}
\startdata
\#1 & 1.65$^{+1.25}_{-1.25}$ & 30 \\
\#2 & 1.83$^{+1.27}_{-1.28}$ & 30 \\
\#3 & 1.85$\pm$1.55  & 30 \\
\#4 & 1.58$\pm$0.98  & 30 \\
\#5 & 1.93$^{+1.27}_{-1.28}$  & 30 \\  
\#6 & 1.92$\pm$1.69  & 30 \\
\#7 & 2.80$\pm$1.60  & 40 
\enddata
\end{deluxetable}
\clearpage

\begin{deluxetable}{lccc}
\tablecaption{Cutoff Frequencies and Cutoff Energies of Cosmic-Ray Electrons 
Accelerated by Galactic SNRs As Derived from the {\it SRCUT} Model\tablenotemark{a} 
\label{Table11}}
\tablewidth{0pt}
\tablehead{
& \colhead{Cutoff} & \colhead{Cutoff} \\
& \colhead{Frequency} & \colhead{Energy} \\
& \colhead{$\nu$$_{\rm{cutoff}}$} & \colhead{$E$$_{\rm{cutoff}}$} \\
\colhead{SNR} & \colhead{(Hz)} & \colhead{(TeV)} & 
\colhead{References} 
}
\startdata
CTB 37B & 3.6$\times$10$^{18}$ & 150 & (1) \\
Kes 73\tablenotemark{b} & 1.5$\times$10$^{18}$ & 100 & (2) \\
G1.9$+$0.3 & 1.4$\times$10$^{18}$ & 90 & (3) \\
Cas A\tablenotemark{c} & 9.0$\times$10$^{17}$ & 80 & (4) \\
& 3.2$\times$10$^{17}$ & 50 & (2) \\
G28.6$-$0.1\tablenotemark{d} & 7.0$\times$10$^{17}$ & 70 & (4) \\
G330.2$+$1.0\tablenotemark{e} & 3.3$\times$10$^{17}$ & 50 & (6) \\
G266.2$-$1.2\tablenotemark{f} & 2.8$\times$10$^{17}$ & 40 & (7) \\
G156.2$+$5.7 & 2.4$\times$10$^{17}$ & 40 & (8) \\
                          & 1.5$\times$10$^{16}$ & 9.7 & (9) \\
G347.3$-$0.5 & 6.3$\times$10$^{17}$ & 60 & (10) \\
& 2.3$\times$10$^{17}$ & 40 & (11) \\
& 2.2$\times$10$^{17}$ & 40 & (12) \\
SN 1006\tablenotemark{c} & 1.1$\times$10$^{17}$ & 30 & (13) \\ 
& 6.0$\times$10$^{16}$ & 20 & (14) \\
Kepler & 1.1$\times$10$^{17}$ & 30 & (2) \\
Tycho & 8.8$\times$10$^{16}$ & 20 & (2) \\
RCW 86 & 8.5$\times$10$^{16}$ & 20 & (15) \\ 
G352.7$-$0.1 & 6.6$\times$10$^{16}$ & 20 & (2) \\
3C 397 & 3.4$\times$10$^{16}$ & 20 & (2)  \\
W49B & 2.4$\times$10$^{16}$ & 10 & (2) \\
G349.7$+$0.2 & 1.8$\times$10$^{16}$ & 10 & (2) \\
3C 396 & 1.6$\times$10$^{16}$ & 10 & (2) \\
G346.6$-$0.2 & 1.5$\times$10$^{16}$ & 9.7 & (2) \\
3C 391 & 1.4$\times$10$^{16}$ & 9.4 & (2) \\
G11.2$-$0.3\tablenotemark{b} (SN 386) & 1.2$\times$10$^{16}$ & 8.7 & (2) \\
RCW 103\tablenotemark{b} & 1.2$\times$10$^{16}$ & 8.7 & (2) 
\enddata
\tablerefs{(1) \citet{Nakamura09}, 
(2) \citet{Reynolds99}, (3) \citet{Reynolds08}, (4) \citet{Stage06},
(5) \citet{Ueno03}, (6) \citet{Park09},  (7) This paper, (8) \citet{Pannuti04}, (9) \citet{Katsuda09b},
(10) \citet{Lazendic04}, (11) \citet{Takahashi08}, (12) \citet{Pannuti03}, 
(13) \citet{Allen08b}, (14) \citet{Reynolds96}, (15) \citet{Rho02}.}
\tablenotetext{a}{Using the published values for $\nu$$_{\rm{cutoff}}$ as provided by the
references and listed here, we have calculated $E$$_{\rm{cutoff}}$ assuming a magnetic field 
strength B=10 $\mu$G.}
\tablenotetext{b} {\citet{Reynolds99}
noticed the presence of a central hard X-ray source in these SNRs which may not
be well resolved from the hard X-ray flux observed from the entire SNR due to the
lower angular resolution capabilities of {\it ASCA.}}
\tablenotetext{c}{In the cases of Cas A and SN 1006, estimates of $\nu$$_{\rm{cutoff}}$
have been presented in the literature based on integrated spectra obtained from
{\it ASCA} observations (see \citet{Reynolds99} and \citet{Reynolds96}, respectively)
as well as {\it Chandra} observations, where the superior angular resolution 
capabilities have revealed a range of spatially-dependent values (see \citet{Stage06} and
\citet{Allen08b}, respectively. For the sake of completeness, we list both the value
derived from the {\it ASCA} observation and the maximum value derived from the
{\it Chandra} observation.}
\tablenotetext{d}{\citet{Ueno03} calculated a magnetic field strength of $B$ = 8 $\mu$G
for this SNR based on equipartition arguments.}
\tablenotetext{e}{\citet{Park09} calculated this cutoff energy based on an 
estimated downstream magnetic field strength of $B$ = 10 -- 50 $\mu$G.}
\tablenotetext{f}{The value for $\nu$$_{\rm{cutoff}}$ presented here corresponds to
the maximum value derived from fits to spectra extracted from the different regions of
the northwestern rim complex of G266.2$-$1.2. Specifically, this maximum value was
obtained for the fit to the extracted spectrum of Region \#7 (see Section 
\ref{NonThermalSubSection} and Table \ref{Table8}.)}
\end{deluxetable}
\clearpage

\begin{deluxetable}{lcc}
\tablecaption{Cutoff Frequencies and Cutoff Energies of
Cosmic-Ray Electrons Accelerated by Large Magellanic Cloud SNRs As Derived From  
the {\it SRCUT} Model\tablenotemark{a} \label{Table12}}
\tablewidth{0pt}
\tablehead{
& \colhead{Cutoff} & \colhead{Cutoff} \\
& \colhead{Frequency} & \colhead{Energy} \\
& \colhead{$\nu$$_{\rm{cutoff}}$} & \colhead{$E$$_{\rm{cutoff}}$} \\
\colhead{SNR} & \colhead{(Hz)} & \colhead{(TeV)} 
}
\startdata
DEM L71 & 2.49$\times$10$^{17}$ & 40 \\
N49B & 6.76$\times$10$^{16}$ & 20 \\
N103B & 4.35$\times$10$^{16}$ & 20 \\
N23 & 4.11$\times$10$^{16}$ & 20 \\
N132D & 3.14$\times$10$^{16}$ & 10 \\
0509$-$67.5 & 2.90$\times$10$^{16}$ & 10 \\
0548$-$70.4 & 1.93$\times$10$^{16}$ & 10 \\
N63A & 1.93$\times$10$^{16}$ & 10 \\
0534$-$69.9 & 2.42$\times$10$^{15}$ & 3.9 \\
0519$-$69.0 & 1.45$\times$10$^{15}$ & 3.0 \\
0453$-$68.5 & 9.66$\times$10$^{14}$ & 2.5
\enddata 
\tablenotetext{a}{All of the listed values for $\nu$$_{\rm{cutoff}}$ have been
taken from \citet{Hendrick01}. Using these values, we have calculated 
$E$$_{\rm{cutoff}}$ assuming a magnetic field strength B=10 $\mu$G.}
\end{deluxetable}

\begin{thebibliography}{}
\bibitem[Acero et al.(2010)]{Acero10} Acero, F., Aharonian, F. A., Akhperjanian, A. G., 
   Anton, G., Barres de Almeida, U., Bazer-Bachi, A. R., Becherini, Y., Behera, B., Beilicke, M.,
    Bernl\"{o}hr, K., Bochow, A., Boisson, C., et al. 2010, \aap, 516, 62
\bibitem[Aharonian et al.(2004)]{Aharonian04} Aharonian, F. A., Akhperjanian, A. G.,
    Aye, K.-M., Bazer-Bachi, A. R., Beilicke, M., Benbow, W., Berge, D., Berghaus, P.,
    Bernl\"{o}hr, K., Bolz, O., Boisson, C., Borgmeier, C., et al. 2004, Nature, 432, 75
\bibitem[Aharonian et al.(2005)]{Aharonian05} Aharonian, F. A., Akhperjanian, A. G., 
    Bazer-Bachi, A. R., Beilicke, M., Benbow, W., Berge, D., Bernl\"{o}hr, K., Boisson, C.,
    Bolz, O., Borrel, V., Braun, I., Breitling, F., et al. 2005, \aap, 437, L7 
\bibitem[Aharonian et al.(2006)]{Aharonian06} Aharonian, F. A., Akhperjanian, A. G., 
   Bazer-Bachi, A. R., Beilicke, M., Benbow, W., Berge, D., Bernl\"{o}hr, K., Boisson, C.,
    Bolz, O., Borrel, V., Braun, I., Breitling, F., et al. 2006, \aap, 449, 223
\bibitem[Aharonian et al.(2007a)]{Aharonian07a} Aharonian, F. A., Akhperjanian, A. G., 
    Bazer-Bachi, A. R., Beilicke, M., Benbow, W., Berge, D., Bernl\"{o}hr, K., Boisson, C., 
     Bolz, O, Borrel, V., Braun, I., Brown, A. M., et al., 2007a, \apj,  661, 236
\bibitem[Aharonian et al.(2007b)]{Aharonian07b} Aharonian, F. A., Akhperjanian, A. G.,
     Bazer-Bachi, A. R., Beilicke, M., Benbow, W., Berge, D., Bernl\"{o}hr, K, Boisson, C., 
     Bolz, O., Borrel, V., Braun, I., Brion, E., et al. 2007b, \aap, 464, 235
\bibitem[Allen et al.(1997)]{Allen97} Allen, G. E., Keohane, J. W., 
    Gotthelf, E. V., Petre, R., Jahoda, K., Rothschild, R. E., Lingenfelter,
    R. E., Heindl, W. A., Marsden, D., Gruber, D. E., Pelling, M. R. \&
    Blanco, P. R. 1997, \apj, 487, L97
\bibitem[Allen et al.(2001)]{Allen01} Allen, G. E., Petre, R. \& Gotthelf,
    E. V. 2001, \apj, 558, 739
\bibitem[Allen et al.(2008a)]{Allen08a} Allen, G.~E., Stage, 
M.~D., \& Houck, J.~C.\ 2008a, International Cosmic Ray Conference, 2, 839 
\bibitem[Allen et al.(2008b)]{Allen08b} Allen, G. E., Houck, J. C. \& Sturner, S. J. 2008b, 
    \apj, 683, 773
\bibitem[Aschenbach et al.(1995)]{Aschenbach95} Aschenbach, B., Egger, R. \&
    Tr\"{u}mper, J. 1995, Nature, 373, 587
\bibitem[Arnaud(1996)]{Arnaud96} Arnaud, K. A. 1996, Astronomical Data Analysis Software
    and Systems V, eds. Jacoby, G. and Barnes, J., p17, ASP Conf. Series Volume 101.
\bibitem[Aschenbach(1998)]{Aschenbach98} Aschenbach, B. 1998, 
    Nature, 396, 141
\bibitem[Aschenbach et al.(1999)]{Aschenbach99} Aschenbach, B.,
    Iyudin, A. F. \& Sch\"{o}nfelder, V. 1999, \aap, 350, 997
\bibitem[Aschenbach(2002)]{Aschenbach02} Aschenbach, B. 2002, Proceedings of
    the 270th WE-Heraeus Seminar on Neutron Stars, Pulsars and Supernova
    Remnants, Physikzentrum Bad Honnef, eds. W. Becker, H. Lesch \&
     J. Tr\"{u}mper, MPE Report 278, 18 (astro-ph/0208492)
\bibitem[Bamba et al.(2001)]{Bamba01} Bamba, A., Ueno, M., Koyama, K.
    \& Yamauchi, S. 2001, \pasj, 53, L21
\bibitem[Bamba et al.(2003)]{Bamba03} Bamba, A., Yamazaki, R., Ueno, M.
    \& Koyama, K. 2003, \apj, 589, 827
\bibitem[Bamba(2004)]{Bamba04} Bamba, A. 2004, Ph.D Thesis, Kyoto University
\bibitem[Bamba et al.(2005a)]{Bamba05a} Bamba, A., Yamazaki, R. \& Hiraga, J. S. 2005a,
    \apj, 632, 294
\bibitem[Bamba et al.(2005b)]{Bamba05b} Bamba, A., Yamazaki, R., Yoshida, T., Terasawa, T.
    \& Koyama, K. 2005b, \apj, 621, 793
\bibitem[Bamba et al.(2008)]{Bamba08} Bamba, A., Fukazawa, Y., Hiraga, J. S., Hughes, J. P.,
    Katagiri, H., Kokubun, M., Koyama, K., Miyata, E., Mizuno, T., Mori, K., Nakajima, H., Ozaki, M.,
    et al. 2008, \pasj, 60, 153 
\bibitem[Bell(1987)]{Bell87} Bell, A. R. 1987, \mnras, 225, 615
\bibitem[Benbow(2005)]{Benbow05} Benbow, W., for the HESS Collaboration, 
    in proceedings of the Gamma 2004 Symposium on High-Energy Gamma-Ray Astronomy,
    Vol. 745 (AIP Conference Proceedings, eds. F.A. Aharonian, H. J. V"{o}lk and D. Horns), 611
\bibitem[Berezhko \& Ellison(1998)]{Berezhko99} Berezhko, E. G. \& Ellison, D. C. 1999, 
    \apj, 526, 389
\bibitem[Berezhko \& V\"{o}lk(2008)]{Berezhko08} Berezhko, E. G. \& V\"{o}lk, H. J. 2008, \aap, 
    492, 695
\bibitem[Borkowski et al.(2001)]{Borkowski01} Borkowski, K. J., Rho, J., Reynolds, S. P. \& 
Dyer, K. K. 2001, \apj, 550, 334 
\bibitem[Burgess \& Zuber(2000)]{Burgess00} Burgess, C. P. \& Zuber, K.
    2000, Astroparticle Physics, 14, 1
\bibitem[Cassam-Chena\"{i} et al.(2004)]{CassamChenai04} Cassam-Chena\"{i},
    G., Decourchelle, A., Ballet, J., Sauvageot, J.-L., Dubner, G. \&
    Giacani, E. 2004, \aap, 427, 199
\bibitem[Cassam-Chena\"{i} et al.(2008)]{CassamChenai08} Cassam-Chena\"{i}, G., 
    Hughes, J. P., Reynoso, E. M., Badenes, C. \& Moffett, D. 2008, \apj, 680, 1180
\bibitem[Cha et al.(1999)]{Cha99} Cha, A. N., Sembach, K. R. \& Danks, A. C. 1999, \apj, 515, L25
\bibitem[Chen \& Gehrels(1999)]{Chen99} Chen, W. \& Gehrels, N. 1999,
    \apj, 514, L103
\bibitem[Combi et al.(1999)]{Combi99} Combi, J. A., Romero, G. E. \&
    Benaglia, P. 1999, \apj, 519, L177
\bibitem[Duncan et al.(1995)]{Duncan95} Duncan, A. R., Stewart, R. T., 
    Haynes, R. F. \& Jones, K. L. 1995, \mnras, 277, 36
\bibitem[Duncan \& Green(2000)]{Duncan00} Duncan, A. R. \& Green, D. A.,
    2000, \aap, 364, 732
\bibitem[Dyer et al.(2001)]{Dyer01} Dyer, K. K., Reynolds, S. P., Borkowski,
     K. J., Allen, G. E. \& Petre, R. 2001, \apj, 551, 439
\bibitem[Dyer et al.(2004)]{Dyer04} Dyer, K. K., Reynolds, S. P. \& 
     Borkowski, K. J. 2004, \apj, 600, 752
\bibitem[Ellison \& Reynolds(1991)]{Ellison91} Ellison, D. C. \& Reynolds, S. P. 1991,
     \apj, 382, 242
\bibitem[Ellison et al.(2001)]{Ellison01} Ellison, D. C., Slane, P. O. \& Gaensler, B. M. 2001, 
     \apj, 563, 191
\bibitem[Ellison \& Vladimirov(2008)]{Ellison08} Ellison, D. C. \& Vladimirov, A. 2008, \apj, 
673, L47
\bibitem[Enomoto et al.(2002)]{Enomoto02} Enomoto, R., Tanimori, T., Naito, T.,
    Yoshida, T., Yanagita, S., Mori, M., Edwards, P. G., Asahara, A., Bicknell, G. V.,
     Gunji, S., Hara, S., Hara, T. et al., 2002, Nature, 416, 823 
\bibitem[Fang et al.(2009)]{Fang09} Fang, J., Zhang, L., Zhang, J. F., Tang, Y.Y. \& Yu, H. 2009,
    \mnras, 392, 925
\bibitem[Filipovi\'{c} et al.(2001)]{Filipovic01} Filipovi\'{c}, M. D., Jones, P. A. \& Aschenbach, B.
    2001, ``Young Supernova Remnants: Eleventh Astrophysics Conference," AIP Conference
    Proceedings, Volume 565, pp. 267 
\bibitem[Fruscione et al.(2006)]{Fruscione06} Fruscione, A. et al. 2006, Proc. SPIE, 6270, 60
\bibitem[Fukui et al.(2003)]{Fukui03} Fukui, Y., Moriguchi, Y., Tamira, K.,
     Yamamoto, H., Tawara, Y., Mizuno, N., Onishi, T., Mizuno, A., 
     Uchiyama, Y., Hiraga, J., Takahashi, T., Yamashita, K. \& Ikeuchi, S.
     2003, \pasj, 55, L61 
\bibitem[Garmire et al.(2003)]{Garmire03} Garmire, G. P., Bautz, M. W., Ford, P. G., Nousek. J. A.
    \& Ricker, Jr., G. R.  2003, SPIE, 4851, 28
\bibitem[Gooch(2006)]{Gooch06} Gooch, R., 2006, KARMA Users Manual, ATNF
\bibitem[Gotthelf et al.(2001)]{Gotthelf01} Gotthelf, E. V., Koralesky, B., 
    Rudnick, L., Jones, T. W., Hwang, U. \& Petre, R. 2001, \apj, 552, L39
\bibitem[Green \& Stephenson(2004)]{Green04} Green, D. A. \& Stephenson, F. R.
    2004, Astroparticle Physics, 20, 613
    \bibitem[Green(2009a)]{Green09a} Green, D. A., 2009a, Bulletin of the Astronomical
    Society of India, 37, 45.
\bibitem[Green(2009b)]{Green09b} Green, D. A., 2009b, `A Catalogue of Galactic 
   Supernova Remnants (2009 March version)', Astrophysics Group, Cavendish Laboratory, 
   Cambridge, United Kingdom (available at
   ``http://www.mrao.cam.ac.uk/surveys/snrs/").
\bibitem[Griffith \& Wright(1993)]{Griffith93}Griffith, M. R. \& Wright, A. E. 
   1993, \aj, 105, 1666 
\bibitem[Hara et al.(1993)]{Hara93} Hara, T. et al. 1993, Nucl. Instr. Meth. Phys. Res. A 332, 300
\bibitem[Helder \& Vink(2008)]{Helder08} Helder, E. A. \& Vink, J. 2008, \apj, 686, 1094
\bibitem[Hendrick \& Reynolds(2001)]{Hendrick01} Hendrick, S. P. \& 
    Reynolds, S. P. 2001, \apj, 559, 903
\bibitem[Hiraga et al.(2005)]{Hiraga05}Hiraga, J. S., Uchiyama, Y., Takahashi, T. 
   \&  Aharonian, F. A. 2005, \aap, 431, 953
\bibitem[Hiraga et al.(2009)]{Hiraga09}Hiraga, J. S., Kobayashi, Y., Tamagawa, T., Hayato, A., 
    Bamba, A., Terada, Y., Petre, R., Katagiri, H. \& Tsunemi, H. 2009, \pasj, 61, 275 
\bibitem[Holt et al.(1994)]{Holt94} Holt, S. S., Gotthelf, E. V., 
    Tsunemi, H. \& Negoro, H. 1994, \pasj, 46, L151
\bibitem[Huang et al.(2007)]{Huang07} Huang, C.-Y., Park, S.-E., Pohl, M. \& Daniels,
    C. D. 2007, Astroparticle Physics, 27, 429
\bibitem[Hughes(2000)]{Hughes00} Hughes, J. P. 2000, \apj, 545, L53
\bibitem[Hwang et al.(2000)]{Hwang00} Hwang, U., Holt, S. S.  \& Petre, R. 
    2000, \apj, 537, L118
\bibitem[Hwang et al.(2002)]{Hwang02} Hwang, U., Decourchelle, A., Holt, S. S.
    \& Petre, R. 2002, \apj, 581, 1101
\bibitem[Iyudin et al.(1998)]{Iyudin98} Iyudin, A. F., Sch\"{o}nfelder, V.,
    Bennett, K., Bloemen, H., Diehl, R., Hermsen, W., Lichti, G. G., 
    van der Meulen, R. D., Ryan, J. \& Winkler, C. 1998, Nature, 396, 142
\bibitem[Iyudin et al.(2005)]{Iyudin05} Iyudin, A. F., Aschenbach, B., Becker,
    W., Dennerl, K. \& Haberl, F. 2005, \aap, 429, 225 
\bibitem[Jones et al.(2003)]{Jones03} Jones, T. J., Rudnick, L., DeLaney, T. \& Bowden, J. 
    2003, \apj, 587,227
\bibitem[Katagiri et al.(2005)]{Katagiri05} Katagiri, H., Enomoto, R., Ksenofontov, L. T.,
    Mori, M., Adachi, Y., Asahara, A., Bicknell, G. V., Clay, R. W., Doi, Y., Edwards, P. G.,
    Gunji, S., Hara, S. et al. 2005, \apj, 619, L163
\bibitem[Katsuda et al.(2008)]{Katsuda08} Katsuda, S., Tsunemi, H. \& Mori, K., 2008,
    \apj, 678, L35
\bibitem[Katsuda et al.(2009a)]{Katsuda09a} Katsuda, S., Petre, R., Long, K. S., Reynolds, S. P.,
   Winkler, P. F., Mori, K. \& Tsunemi, H. 2009a, \apj, 692, L105 
\bibitem[Katsuda et al.(2009b)]{Katsuda09b} Katsuda, S., Petre, R., Hwang, U., Yamaguchi, H.,
    Mori, K. and Tsunemi, H. 2009b, \pasj, 61, S155
\bibitem[Katz \& Waxman(2008)]{Katz08} Katz, B. \& Waxman, E. 2008, \jcap, 1, 18
\bibitem[Kawachi et al.(2001)]{Kawachi01} Kawachi, A., Hayami, Y., Jimbo, J., 
    Kamei, S., Kifune, T., Kubo, H., Kushida, J., LeBohec, S., Miyawaki, K.,
    Mori, M., Nishijima, K., Patterson, J. R., et al. 2001, Astroparticle Physics, 14, 261  
\bibitem[Keohane(1998)]{Keohane98} Keohane, J. W. 1998, Ph.D Thesis,
    University of Minnesota
 \bibitem[Ksenofontov et al.(2010)]{Ksenofontov10} Ksenofontov, L. T., V\"{o}lk, H. J., 
    Berezhko, E. G. 2010, \apj, 714, 1187
\bibitem[Koyama et al.(1995)]{Koyama95} Koyama, K., Petre, R., Gotthelf, E. V., 
    Hwang, U., Matsuura, M., Ozaki, M. \& Holt, S. S. 1995, Nature,
    378, 255
\bibitem[Koyama et al.(1997)]{Koyama97} Koyama, K., Kinugasa, K., 
    Matsuzaki, K., Nishiuchi, M., Sugizaki, M., Torii, K., Yamauchi,
    S. \& Aschenbach, B. 1997, \pasj, 49, L7
\bibitem[Koyama et al.(2001)]{Koyama01} Koyama, K., Ueno, M., Bamba, A.
    \& Ebisawa, K. 2001, Proceedings of
    the Symposium ``New Visions of the X-ray Universe in the XMM-Newton
    and Chandra Era," Noordwijk, The Netherlands (arXiv: astro-ph/0202009)
\bibitem[Lazendic et al.(2004)]{Lazendic04} Lazendic, J. S., Slane, P. O., 
    Gaensler, B. M., Reynolds, S. P., Plucinsky, P. P. and Hughes, J. P.
    2004, \apj, 602, 271
\bibitem[Long et al.(2003)]{Long03} Long, K. S., Reynolds, S. P., Raymond,
     J. C., Winkler, P. F., Dyer, K. K. \& Petre, R. 2003, \apj, 
     586, 1162
\bibitem[Longair(1994)]{Longair94}Longair, M. S. (1994), High Energy Astrophysics,
     Volume 2, Stars, the Galaxy and the Interstellar Medium, Cambridge University Press,
     Cambridge, UK. 
\bibitem[Lu \& Aschenbach(2000)]{Lu00} Lu, F. J. \& Aschenbach, B. 2000,
     \aap, 362, 1083
\bibitem[Lyutikov \& Pohl(2004)]{Lyutikov04}Lyutikov, M. \& Pohl, M. 2004, 
     \apj, 609. 785
\bibitem[Maeda et al.(2009)]{Maeda09}Maeda, Y., Uchiyama, Y., Bamba, A., Kosugi, H., 
    Tsunemi, H., Helder, E. A., Vink, J., Kodaka, N., Terada, Y., Fukazawa, Y., Hiraga, J., 
    Hughes, J. P.,
    Kokubun, M., Kouzu, T., Matsumoto, H., Miyata, E., Nakamura, R., Okada, S., Someya, K., 
    Tamagawa, T., Tamura, K., Totsuka, K., Tsuboi, Y., Ezoe, Y., Holt, S. S., Ishida, M., Kamae, T., 
    Petre, R. \& Takahashi, T. 2008, \pasj, 61, 1217
\bibitem[Malkov et al.(2005)]{Malkov05} Malkov, M. A., Diamond, P. H. \& Sagdeev,
     R. Z. 2005, \apj, 624, L37
\bibitem[May et al.(1988)]{May88} May, J., Murphy, D. C. \& Thaddeus, P.
     1988, \aaps, 73, 51
\bibitem[McClure-Griffiths et al.(2001)]{McClure01} McClure-Griffiths, N. M., Green, A. J.,
     Dickey, J. M., Gaensler, B. M., Haynes, R. F. \& Wieringa, M. H. 2001, \apj, 551, 394
\bibitem[Mereghetti(2001)]{Mereghetti01} Mereghetti, S. 2001, \apj, 548,  L213
\bibitem[Moraitis \& Mastichiadis(2007)]{Moraitis07} Moraitis, K. \& Mastichiadis, A. 2007,
    \aap, 462, 173
\bibitem[Morlino et al.(2009)]{Morlino09} Morlino, G., Amato, E. \& Blasi, P. 2009, \mnras, 392, 240
\bibitem[Motizuki et al.(2009)]{Motizuki09} Motizuki, Y., Takahashi, K., Makishima, K., Bamba, A.,
    Nakai, Y., Yano, Y., Igarashi, M., Motoyama, H., Kamiyama, K., Suzuki, K. \& Imamura, T. 2009,
    Nature, submitted (http://arxiv.org/abs/0902.3446)
\bibitem[Muraishi et al.(2000)]{Muraishi00} Muraishi, H., Tanimori, T., 
    Yanagita, S., Yoshida, T., Moriya, M., Kifune, T., Dazeley, S. A., 
    Edwards, P. G., Gunji, S., Hara, S.,
    Hara, T., Kawachi, A., et al. 2000, \aap, 354, L57 
\bibitem[Murphy(1985)]{Murphy85} Murphy, D. C. 1985, Ph.D Thesis,
    Massachusetts Institute of Technology
\bibitem[Nakamura et al.(2009)]{Nakamura09} Nakamura, R., Bamba, A., Ishida, M., 
    Nakajima, H., Yamazaki, R., Terada, Y., P\"uhlhofer, G. \& Wagner, S. J. 2009, \pasj, 61, 197
\bibitem[Ogasawara et al.(2007)]{Ogasawara07} Ogasawara, T., Yoshida, T.,
   Yamagita, S. \& Kifune, T. 2007, \apss, 309, 401
\bibitem[Pannuti et al.(2003)]{Pannuti03} Pannuti, T. G., Allen, G. E.,
    Houck, J. C. \& Sturner, S. J. 2003, \apj, 593, 377
\bibitem[Pannuti \& Allen(2004)]{Pannuti04} Pannuti, T. G. \& Allen, G. E. 
    2004, AdSpR, 33, 434
\bibitem[Park et al.(2009)]{Park09} Park, S., Kargaltsev, O., Pavlov, G. G., Mori, K., 
    Slane, P. O., Hughes, J. P., Burrows, D. N. and Garmire, G. P.  2009, ApJ, 695, 431
\bibitem[Pavlov et al.(2001)]{Pavlov01} Pavlov, G. G., Sanwal, D., 
    Kiziltan, B. \& Garmire, G. P. 2001, \apj, 559, L131
\bibitem[Pellizzoni et al.(2002)]{Pellizzoni02} Pellizzoni, A., 
    Mereghetti, S. \& De Luca, A. 2002, \aap, 393, L65
\bibitem[Plaga(2008)]{Plaga08} Plaga, R. 2008, New Astronomy, 13, 73
\bibitem[Pohl et al.(2005)]{Pohl05} Pohl, M., Yan, H. \& Lazarian, A. 2005, \apj,
    626, L101
\bibitem[Redman et al.(2000)]{Redman00} Redman, M. P., Meaburn, J., 
    O'Connor, J. A., Holloway, A. J. \& Bryce, M. 2000, \apj, 543, L153
\bibitem[Redman et al.(2002)]{Redman02} Redman, M. P., Meaburn, J., 
    Bryce, M., Harman, D. J. \& O'Brien, T. J. 2002, \mnras, 336, 1093
\bibitem[Redman \& Meaburn(2005)]{Redman05} Redman, M. P. \& Meaburn, J.
    2005,\mnras, 356, 969
\bibitem[Reed et al.(1995)]{Reed95} Reed, J. E., Hester, J. J., Fabian, A. C. 
    and Winkler, P. F. 1995, \apj, 440, 706
\bibitem[Reimer \& Pohl(2002)]{Reimer02} Reimer, O. \& Pohl, M. 2002, \aap, 390, L43
\bibitem[Reynolds(1996)]{Reynolds96} Reynolds, S. P. 1996, \apj, 459, L13
\bibitem[Reynolds(1998)]{Reynolds98} Reynolds, S. P. 1998, \apj,
    493, 375
\bibitem[Reynolds \& Keohane(1999)]{Reynolds99} Reynolds, S. P. \& 
    Keohane, J. W. 1999, \apj, 525, 368
\bibitem[Reynolds et al.(2007)]{Reynolds07} Reynolds, S. P., Borkowski, K. J., Hwang, U.,
Hughes, J. P., Badenes, C., Laming, J. M. \& Blondin, J. M. 2007, \apjl, 668, L135
\bibitem[Reynolds et al.(2008)]{Reynolds08} Reynolds, S. P., Borkowski, K. J.,
    Green, D. A., Hwang, U., Harrus, I. \& Petre, R. 2008, \apj, 680, L41
\bibitem[Reynolds et al.(2009)]{Reynolds09} Reynolds, S. P., Borkowski, K. J., Green, D. A., 
    Hwang, U., Harrus, I. \& Petre, R.  2009, \apj, 695, L149
\bibitem[Reynoso \& Goss(1999)]{Reynoso99} Reynoso, E. M. \& Goss, W. M. 1999, \aj, 118, 926
\bibitem[Rosado et al.(1996)]{Rosado96} Rosado, M., Ambrocio-Cruz, P., Le Coarer, E. \& 
     Marcelin, M. 1996, \aap, 315, 243
\bibitem[Rho et al.(2002)]{Rho02} Rho, J., Dyer, K. K., Borkowski, K. J.
    \& Reynolds, S.P. 2002, \apj, 581, 1116
\bibitem[Rho et al.(2003)]{Rho03} Rho, J., Reynolds, S. P., Reach, W. T., 
    Jarrett, T. H., Allen, G. E. \& Wilson, J. C. 2003, \apj, 592, 299
\bibitem[Rood et al.(1979)]{Rood79} Rood, R. T., Sarazin, C. L., Zeller,  
    E. J. \& Parker, B. C. 1979, Nature, 282, 701
\bibitem[Rothenflug et al.(2004)]{Rothenflug04} Rothenflug, R., Ballet., J., 
    Dubner, G., Giacani, E., Decourchelle, A. \& Ferrando, P. 2004, \aap, 425, 
    121
\bibitem[Sankrit et al.(2003)]{Sankrit03} Sankrit, R., Blair, W. P. \&
    Raymond, J. C. 2003, \apj, 589, 242
\bibitem[Sault \& Killeen(2006)]{Sault06} Sault, R. \& Killeen, N. 2006, 
    MIRIAD Users Guide, ATNF
\bibitem[Sch\"{o}nfelder et al.(2000)]{Schoenfelder00}Sch\"{o}nfelder, V., 
    Bloemen, H., Collmar, W., Diehl, R., Hermsen, W., Kn\"{o}dlseder, J.,
    Lichti, G. G., Pl\"{u}schke, S., Ryan, J., Strong, A. \& Winkler,
    C. 2000, in AIP Conf. Proc. 510, Fifth Compton Symposium, ed. M. L.
    McConnell \& J. M. Ryan (New York: AIP), 54 
\bibitem[Slane et al.(1999)]{Slane99} Slane, P., Gaensler, B. M., 
    Dame, T. M., Hughes, J. P., Plucinsky, P. P. \& Green, A. 1999, \apj,
    525, 357 
\bibitem[Slane et al.(2001a)]{Slane01a} Slane, P., Hughes, J. P., 
    Edgar, R. J., Plucinsky, P. P., Miyata, E., Tsunemi, H. \&
    Aschenbach, B. 2001a, \apj, 548, 814
\bibitem[Slane et al.(2001b)]{Slane01b} Slane, P., Hughes, J. P., Edgar, 
    R. J., Plucinsky, P. P., Miyata, E., Tsunemi, H. \& Aschenbach, B.
    2001b, in Holt, S. S. Hwang, U., eds, Young Supernova Remnants, AIP,
    New York, p. 403
\bibitem[Stage et al.(2006)]{Stage06} Stage, M. D., Allen, G. E., Houck, J. C. 
    \& Davis, J. E. 2006, NatPh, 2, 614
\bibitem[Stupar et al.(2005)]{Stupar05} Stupar, M., Filipovi\'{c}, M. D., 
    Jones, P. A. \& Parker, Q. A. 2005, AdSpR, 35, 1047
\bibitem[Sugizaki(1999)]{Sugizaki99} Sugizaki, M. 1999, Ph.D Thesis, 
    University of Tokyo
\bibitem[Takahashi et al.(2008)]{Takahashi08} Takahashi, T., Tanaka, T., Uchiyama, Y.,
    Hiraga, J. S., Nakazawa, K., Watanabe, S., Bamba, A., Hughes, J. P., Katagiri, H., Takaoka, J.,
     Kokubun, M., Koyama, K., Mori, K., Petre, R., Takahashi, H. \& Tsuboi, Y. 2008,
     \pasj, 60, 131
\bibitem[Tanaka et al.(2008)]{Tanaka08} Tanaka, T., Uchiyama, Y., Aharonian, F. A., 
   Takahashi, T., Bamba, A., Hiraga, J. S., Kataoka, J., Kishishita, T., Kokubun, M., 
    Mori, K., Nakazawa, K., Petre, R. et al. 2008, \apj, 685, 988
\bibitem[Tanimori et al.(1998)]{Tanimori98} Tanimori, T., Hayami, Y., Kamei, 
    S., Dazeley, S. A., Edwards, P. G., Gunji, S., Hara, S., Hara, T., 
    Holder, J., Kawachi, A., Kifune, T., Kita, R., et al. 1998, \apj, 497, L25  
\bibitem[Telezhinsky(2009)]{Telezhinsky09} Telezhinsky, I. 2009, APh, 31, 431
\bibitem[Torii et al.(2006)]{Torii06} Torii, K., Uchida, H., Hasuike, K., Tsunemi, H.,
    Yamauchi, Y. \& Shibata, S., \pasj, 58, L11
\bibitem[Tsunemi et al.(2000)]{Tsunemi00} Tsunemi, H., Miyata, E., 
    Aschenbach, B., Hiraga, J. \& Akutsu, D. 2000, \pasj, 52, 887
\bibitem[Turtle et al.(1962)]{Turtle62} Turtle, A. J., Pugh, J. F., Kenderdine, S., Pauliny-Toth,
    I. I. K. 1962, \mnras, 124, 297   
\bibitem[Uchiyama et al.(2003)]{Uchiyama03} Uchiyama, Y., Aharonian, F. A.
     \& Takahashi, T. 2003, \aap, 400, 567
\bibitem[Uchiyama et al.(2007)]{Uchiyama07} Uchiyama, Y., Aharonian, F. A., 
    Tanaka, T., Takahashi, T. \& Maeda, Y. 2007, Nature, 449, 576
\bibitem[Ueno et al.(2003)]{Ueno03} Ueno, M., Bamba, A., Koyama, K. 
     \& Ebisawa, K. 2003, \apj, 588, 338
\bibitem[van der Swaluw \& Achterberg(2004)]{vanderswaluw04} van der Swaluw, E. \&
     Achterberg, A. 2004, \aap, 421, 1021
\bibitem[Vink(2004)]{VinkJ04} Vink, J. 2003, ``The Restless High Energy 
     Universe" Conference Proceedings, Nucl. Physics B. 
     Suppl. Series, eds. E. P. J. van den Heuvel, J. J. M. in 't Zand 
     \& R. A. M. J. Wijers, 132, 21
\bibitem[Vink et al.(2003)]{Vinketal03} Vink, J., Laming, J. M., Gu, M. F., Rasmussen, A. \&
      Kaastra, J. S. 2003, \apj, 587, L31
\bibitem[Vink \& Laming (2003)]{Vink03} Vink, J. \& Laming, J. M. 2003, \apj, 584, 758
\bibitem[Vink(2005)]{Vink05} Vink, J. 2005, ``International Symposium on 
     High-Energy Gamma-Ray Astronomy (Gamma-2004)" Conference Proceedings, 
     Heidelberg, edited by F. A. Aharonian and H. V\"{o}lk (AIP, NY), 745, 160
\bibitem[Vink et al.(2006)]{Vink06} Vink, J., Bleeker, J., van der Heyden, K., Bykov, A., 
     Bamba, A. \& Yamazaki, R. 2006, \apj, 648, L33
\bibitem[Vink(2008)]{Vink08} Vink, J. 2008, \aap, 486, 837
\bibitem[Weisskopf et al.(2002)]{Weisskopf02} Weisskopf, M. C., Brinkman, B.,
     Canizares, C., Garmire, G., Murray, S. \& van Speybroeck, L. P. et al. 
     2002, \pasp, 114, 1
\bibitem[Winkler \& Long(1997)]{Winkler97} Winkler, P. F. \& Long, K. S.
     1997, \apj, 491, 829
\bibitem[Winkler et al.(2003)]{Winkler03} Winkler, P. F., Gupta, G. \& Long, K. S. 2003,
     \apj, 585, 324
\bibitem[Yamaguchi et al.(2004)]{Yamaguchi04}Yamaguchi, H., Ueno, M., Koyama, K.,
Bamba, A. \& Yamauchi, S. 2004, \pasj, 56, 1059
\bibitem[Zirakashvili \& Aharonian(2007)]{Zirakashvili07} Zirakashvili, V. N. \& Aharonian, F.
2007, \aap, 465, 695
\end{thebibliography}
\end{document}